\documentclass[a4paper]{spie}  
\usepackage[]{graphicx}
\usepackage{amssymb,amstext}
\usepackage{cite}
\usepackage{hyperref}

\def\R23{\mbox{$\rm R_{23}$}}

\usepackage{geometry}
\title{High spectral-resolution interferometry down to 1 micron\\ with Asgard/BIFROST at VLTI:\\ Science drivers and project overview
 \footnote{Copyright 2022 Society of Photo-Optical Instrumentation Engineers. One print or electronic copy may be made for personal use only. Systematic reproduction and distribution, duplication of any material in this paper for a fee or for commercial purposes, or modification of the content of the paper are prohibited.}
}

\author{Stefan Kraus\supit{a}, 
Daniel Mortimer\supit{a},
Sorabh Chhabra\supit{a},
Yi Lu\supit{a},
Isabelle Codron\supit{a},\\
Tyler Gardner\supit{a,b},
Narsireddy Anugu\supit{c},
John D.\ Monnier\supit{b},
Jean-Baptiste Le Bouquin\supit{d},\\
Michael Ireland\supit{e},
Frantz Martinache\supit{f},
Denis Defrère\supit{g},
Marc-Antoine Martinod\supit{g}
\skiplinehalf
\supit{a}University of Exeter, School of Physics and Astronomy, Stocker Road, Exeter, UK;\\  
\supit{b}Department of Astronomy, University of Michigan, Ann Arbor, USA; \\
\supit{c}CHARA Array, Mt.\ Wilson Observatory, Mt.\ Wilson, California, USA; \\
\supit{d}Institut de Plan\'etologie et d'Astrophysique de Grenoble, Grenoble, France; \\
\supit{e}Australian National University, Canberra, Australia;\\
\supit{f}Observatoire de la Côte d'Azur, Nice, France; \\
\supit{g}Institute of Astronomy, KU Leuven, 3001 Leuven, Belgium
}

\authorinfo{Further author information: \\
Send correspondence to S.K., E-mail: s.kraus@exeter.ac.uk, Telephone: +44 1392 724125}

\begin{document} 
\maketitle 

\begin{abstract}
We present science cases and instrument design considerations for the BIFROST instrument that will open the
short-wavelength (Y/J/H-band), high spectral dispersion (up to R=25,000) window for the VLT Interferometer. 
BIFROST will be part of the Asgard Suite of instruments and unlock powerful venues for studying 
accretion \& mass-loss processes at the early/late stages of stellar evolution, for detecting accreting 
protoplanets around young stars, and for probing the spin-orbit alignment in directly-imaged planetary 
systems and multiple star systems. Our survey on GAIA binaries aims to provide masses and precision ages
for a thousand stars, providing a legacy data set for improving stellar evolutionary models as well
as for Galactic Archaeology.  
BIFROST will enable off-axis spectroscopy of exoplanets in the 0.025-1” separation range, enabling 
high-SNR, high spectral resolution follow-up of exoplanets detected with ELT and JWST. 
We give an update on the status of the project, outline our key technology choices, and discuss synergies with other instruments in the proposed Asgard Suite of instruments.
\end{abstract}



\section{INTRODUCTION}
\label{sec:intro}

Optical interferometric instruments focussed so far either on the visible waveband (V-band) or wavelengths longwards of 1.5\,$\mu$m (H/K/L/M/N-band), with the intermediary Y and J-band (1.0-1.4\,$\mu$m) receiving much less attention.  However, there are both technical and scientific reasons for targeting these bands, specifically at high spectral resolution:  
On the scientific side, there are unique spectral lines in the YJ-band, including the He\,I 1.083\,$\mu$m accretion-tracing line\cite{fis08} and forbidden lines (e.g. [Fe II] 1.257\,$\mu$m). For hydrogen recombination lines the equivalent width of the lines also typically increases towards shorter wavelengths, with the Paschen series in the YJ-band typically featuring a 5-10$\times$ higher equivalent width than the Brackett series that is accessible in KLM-band.  Line intensities increase further in the Lyman series (ultraviolet) and Balmer series (visible), although these transitions are also stronger affected by poor atmospheric transmission and/or dust extinction.
On the technical side, the YJ-band allows achieving higher angular resolution than the near-/mid-infrared, while state-of-the-art adaptive optics technique enables the high Strehl that is needed to achieve high sensitivity and high precision.  Also, at ESO's Very Large Telescope Interferometer (VLTI), the infrastructure is already equipped for accessing the YJ-band, while this is not the case towards shorter wavelengths.

In this article, we outline our plans for a VLTI instrument that is optimized for the Y/J/H-band (1.05-1.7\,$\mu$m) and for high spectral resolution (up to R=25,000).  The construction of the core capabilities of this instrument, named BIFROST (“Beam-combination Instrument for studying the Formation and fundamental paRameters of Stars and planeTary systems”), are funded through an Consolidator Grant of the European Research Council (``GAIA-BIFROST'', Grant Agreement No.\ 101003096). 

We start with an overview about the key science drivers (section~\ref{sec:science})
and outlined how BIFROST will be integrated in the Asgard Suite of Instruments (section~\ref{sec:asgard}).
Then we outline the BIFROST instrument design \& sub-systems (section~\ref{sec:design}) and 
summarise the planned operating modes (section~\ref{sec:modes}).
We finish with an overview about the project status (section~\ref{sec:conclusions}).

\section{KEY SCIENCE OBJECTIVES}
\label{sec:science}

Below we summarise the key science drivers for BIFROST, as presented at the ``VLT in 2023'' workshop (Kraus et al.\cite{kra19}) and in the recent science whitepaper that we submitted to ESO in March 2022.

\begin{figure}[b]
\centering
\includegraphics[width=1.0\textwidth]{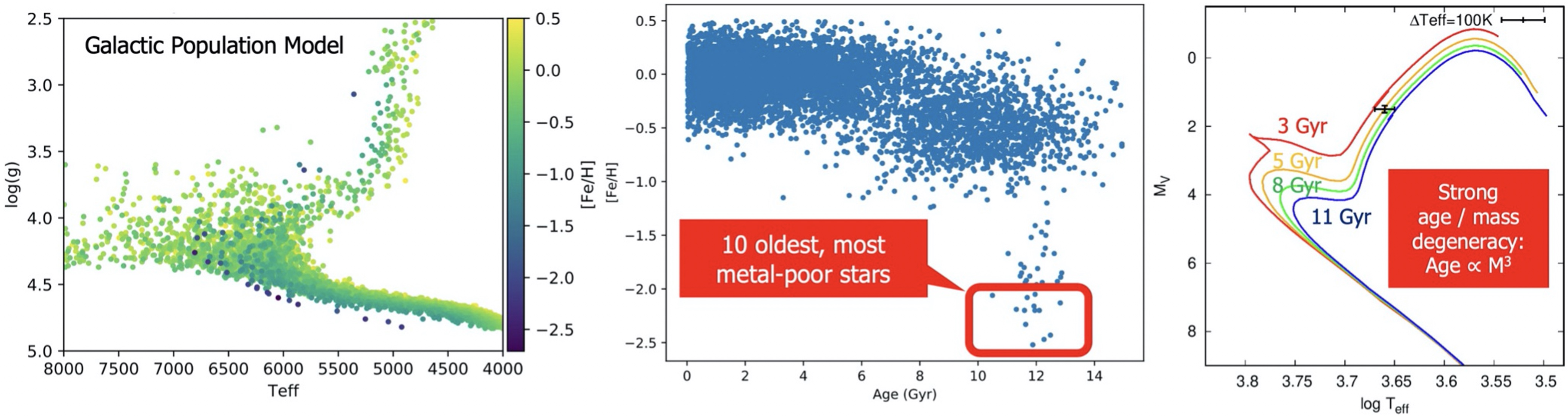} 
\caption{Target sample accessible with BIFROST and the AT, based on a Galactic Population model by Luca Casagrande. Left: Position of the primary stars in the HR diagram – for each star we will be able to obtain a precision dynamical mass. Middle: Metallicity of primaries, plotted as function of stellar age, indicating that our sample will include some of the oldest, most metal-poor stars in the Milky Way that are of particular interest for stellar evolutionary models and Galactic Archaeology (the sample size increases to $\sim50$ with the UTs). Right: Isochrone fitting alone provides only weak constraints on the age of a star due to an age/mass degeneracy– dynamical mass estimates allow us to break this degeneracy. }
\label{fig:gaiasample}
\end{figure}

\subsection{GAIA binaries legacy survey}
\label{sec:sciencegaiabinaries}

Starting with Data Release 3 (DR3), the GAIA mission detects of the order of $\sim28$ million non-single stars, i.e. stars where the astrometric motion indicates the presence of companion(s), including $\sim228,000$ planets\cite{cas08}. GAIA’s RVS instrument also provides precision radial velocities, which is predicted to result in $\sim5$ million systems that are detected both through astrometry and as double-lined spectroscopic binary\cite{gai16}. However, with an angular resolution of 0.1”, GAIA will not be able to spatially resolve most of these systems, but measure instead the astrometric motion of the photocenter between the two stars. As a result, the GAIA sample will face a fundamental degeneracy between the binary flux ratio f and the binary separation $\rho$. For instance, a photocenter motion of 1 mas can be caused by a binary with $f=1:2$ and $\rho=3$\,milliarcseconds (mas), but equally well by a binary with, e.g., $f=1:4$ and $\rho=5$\,mas. This degeneracy introduces uncertainties in the derived binary statistics and prevents dynamical masses ($m_{1}$, $m_{2}$) to be computed (only $m_{23}/(m_{1}+m_{2})^{2}$ can be derived). The degeneracy can be solved for eclipsing systems – however, eclipsing systems are so close that they are likely to have interacted during their evolution, making them unreliable probes for deducing their formation history and for calibrating stellar evolution models.
The proposed GAIA binaries survey will exploit the opportunities that arise from GAIA astrometric+SB2 companion detections and spatially resolve up to $\sim6000$ GAIA binaries in continuum (science case \ref{sec:sciencegaiabinaries}) and, simultaneously, in spectral lines tracing the stellar spin orientations (see science case \ref{sec:sciencespin}). From the continuum observations, we can derive:

\begin{figure}[tb]
\centering
\includegraphics[width=0.9\textwidth]{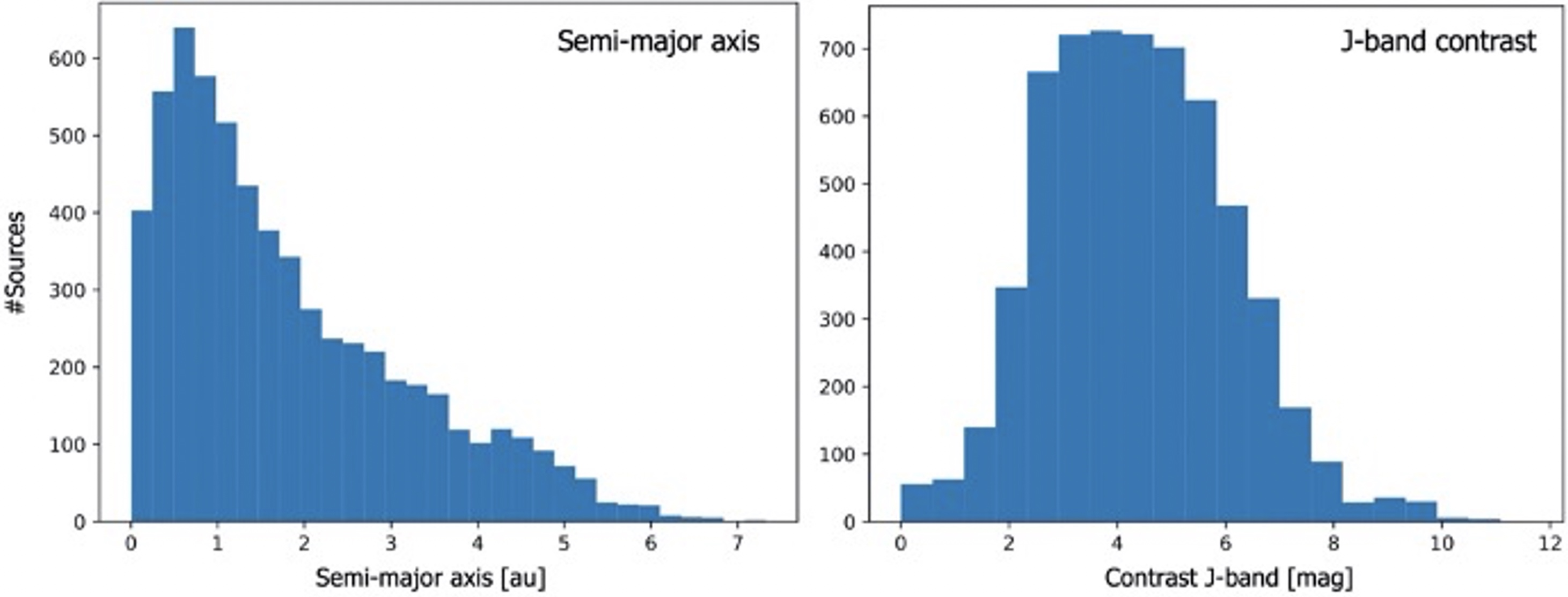} 
\caption{Semi-major axis distribution and brightness contrast distribution for the GAIA binaries that will be accessible with BIFROST and the ATs (derived from GUMS18 population\cite{rob12}).}
\label{fig:semimajoraxis}
\end{figure}

\begin{itemize}
\item {\bf Orbit statistics:} Solving the separation-flux ratio degeneracy inherent to GAIA photocenter measurements will allow us to derive precise semi-major axis and mass ratio statistics. Combining the derived statistics with results from RV/classical-imaging surveys will allow us to search for evidence for a bimodal distribution that could indicate multiple pathways for binary formation, following on tentative findings for a bimodal separation distribution for A stars\cite{duc13} and OB stars\cite{gra18}.

\item {\bf Dynamical masses:} Dynamical masses from binaries constitute the Gold standard for testing and calibrating stellar evolutionary models. These models still face major uncertainties, for instance with respect to the treatment of mass-loss and convective overshooting in massive stars, and the convective mixing length in low-mass stars \cite{fei12,man15}. The uncertainties are particularly notable in the pre-main-sequence (PMS) phase, where predicted and measured masses differ by $\gtrsim 10$\% \cite{sta14}.
We simulate the target sample that will be accessible with BIFROST using two independent stellar population simulations, namely our Galactic Population model and the 2018 “GAIA Universe Model Snapshot” \cite{rob12}. We select sources that produce sufficient photocenter motion and RV modulation to achieve high-SNR GAIA astrometry+RV orbit constraints (SNR>20), have modest binary flux ratios ($\Delta m<5^{\mathrm{m}}$), orbital periods $<10$\,years, are sufficiently wide to be spatially resolved with BIFROST, and are within the sensitivity/observability limits. Both population models predict that BIFROST will be able to measure dynamical masses for up to $\sim 6000$ GAIA binaries spread over the HR diagram (Fig.~\ref{fig:gaiasample}), far exceeding the sample size achieved in any earlier dynamical mass survey. The measured dynamical masses will provide essential input for the refinement of evolutionary models that lie at the foundation of modern astrophysics.

\item {\bf Ages:} For evolved objects (near the Red Giant Branch; $\sim1/4$ of the sample; Fig.~\ref{fig:gaiasample}), the precision masses can be used to derive ages, where we will achieve higher precision ($<10$\%) than competing methods. Covering GAIA binaries out to 1 kpc from the Sun, the proposed survey will allow us to measure the age distribution in our arm of the Milky Way, providing essential input for Galactic Archaeology. This might allow us to uncover regions of episodic star formation or minor merger events that might have shaped the Milky Way in the past \cite{cha20}. Astroseismology offers another method for constraining stellar ages, but relies on a controversial seismic scaling relation that introduces uncertainties up to 15\% on mass \cite{bro18}. The large number of precision masses determined from our survey will provide anchor points to calibrate the scaling relation, thereby increasing the accuracy of the ages that will be derived for ten-thousands of sources with TESS+PLATO \cite{cam16,mig17}.
Importance to the field:  Earlier surveys on orbit statistics or masses needed to observe large samples to identify binary candidates and then monitor the candidates over decades to determine astrometry+RV orbits. Here we will know most orbital elements from GAIA and can select specifically the most interesting systems, where a single BIFROST observation will yield a full-characterized system with dynamical masses. As a result, the survey will provide a legacy data set of orbital parameters, dynamical masses \& precision ages for (literally!) thousand stars, far exceeding earlier work. We will probe systems with periods between 0.1-10 yrs (=separations of a few au; Fig.~\ref{fig:semimajoraxis}) that are particularly valuable, as they fill the gap between AO+RV surveys and correspond to systems that are unlikely to have interacted in the past, providing a pristine sample for stellar evolution studies.
\end{itemize}

 For the ATs, we estimate the accessible sample to be $\sim6000$ GAIA binaries (Fig.~\ref{fig:gaiasample}), where we will record continuum data (LR arm) and line data (HR arm; to determine spin-orbit alignment for science case 3.2) simultaneously. For most systems, it will be possible to derive the information from closure phases and wavelength-differential phases, minimising calibrator star visits, and making the observations highly efficient ($\sim25$ minutes per system). Therefore, the sample size will mainly be limited by the allocated time. The sample can be extended with observations on the UTs to access rare stellar populations, such as very low-mass stars.\newline

{\bf Importance to the field:}  Earlier surveys on orbit statistics or masses needed to observe large samples to identify binary candidates and then monitor the candidates over decades to determine astrometry+RV orbits. Here we will know most orbital elements from GAIA and can select specifically the most interesting systems, where a single BIFROST observation will yield a full-characterized system with dynamical masses. As a result, the survey will provide a legacy data set of orbital parameters, dynamical masses \& precision ages for (literally!) thousand stars, far exceeding earlier work. We will probe systems with periods between 0.1-10 yrs (=separations of a few au; Fig.~\ref{fig:semimajoraxis}) that are particularly valuable, as they fill the gap between AO+RV surveys and correspond to systems that are unlikely to have interacted in the past, providing a pristine sample for stellar evolution studies.\\

\begin{figure}[tb]
\centering
\includegraphics[width=0.9\textwidth]{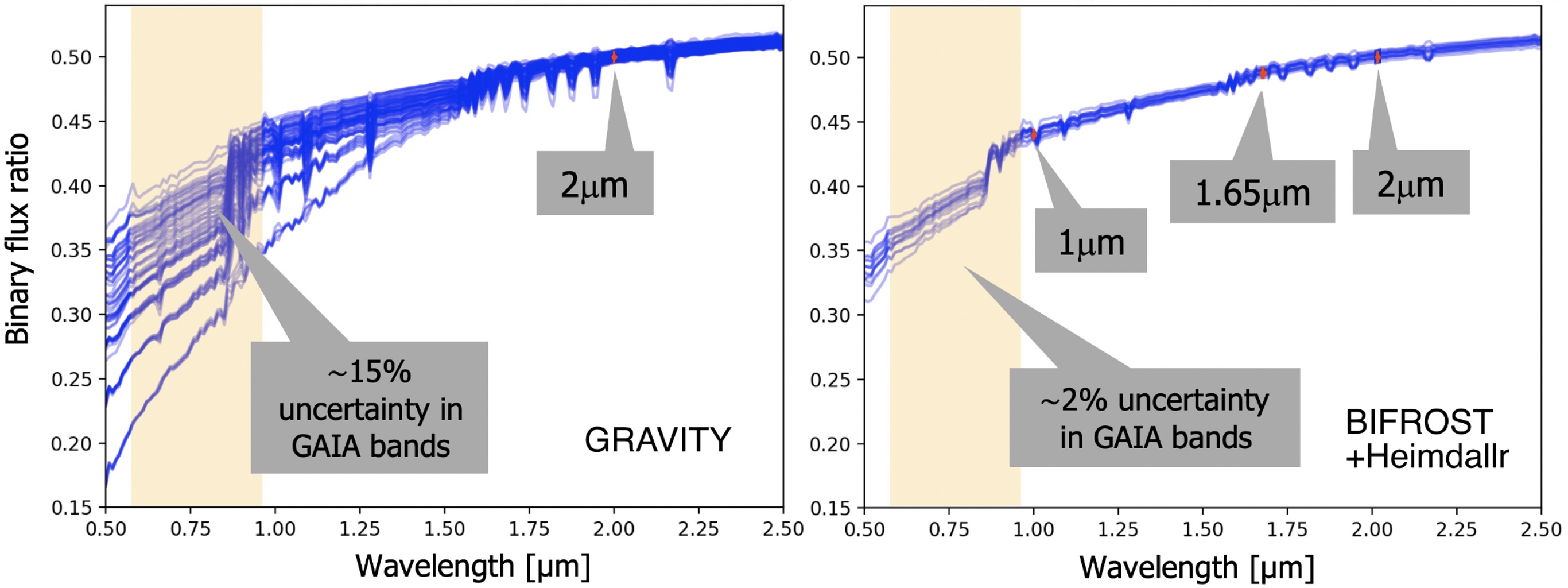} 
\caption{The flux ratio of GAIA binaries needs to be measured close to the GAIA bands (yellow box). Left: Existing VLTI instruments operate at longer wavelengths, introducing 15\% uncertainties when extrapolating to the GAIA bands. Right: Asgard enables 2\% flux-ratio precision.}
\label{fig:gaiafluxratio}
\end{figure}

{\bf Unique capabilities:} We need to measure the binary flux ratio close to the GAIA band (0.4-0.9\,$\mu$m) to derive dynamical masses with good ($\sim3$\%) precision. Existing instruments operate at longer wavelengths and extrapolating the flux ratio towards the GAIA bands induces unacceptably large uncertainties ($\sim15$\% for K-band; Fig.~\ref{fig:gaiafluxratio}).

\subsection{Spin-orbit alignment in binaries and planet-host stars}
\label{sec:sciencespin}

A crucial diagnostic that can tell us about processes involved in the formation and dynamical evolution of binaries and planetary systems is the angle between the stellar rotation axis and the orbital angular momentum vector (spin-orbit alignment, or obliquity). For eclipsing/transiting systems, the sky-projected obliquity can be measured with the Rossiter McLaughlin (RML) effect \cite{que00}. Strong obliquities have been found both for stellar-mass \cite{tri17} and planetary-mass companions \cite{lai11}, incl. 40\% of Hot Jupiters, with half of these on retrograde orbits. However, the obliquity distribution remains largely unexplored for systems with longer periods, where the RML effect cannot be applied. 

We will use BIFROST to measure obliquities for a large sample of GAIA systems with astrometrically-detected companions (from stellar-mass companions down to planetary-mass systems). We estimate that $\sim1300$ of the stars in our GAIA binary sample will have stellar diameters sufficiently large to measure the spin axis orientation with BIFROST, allowing a comparison with the orbit orientation derived from GAIA astrometry. For a subset of binary systems, we can also measure the alignment of the two stellar spin axes (spin-spin alignment). We will include host stars of directly-imaged planets, where the sample size is expected to increase soon with the arrival of JWST and ELT. The resulting obliquity distribution for a large sample of star-star and star-planet systems will allow us to test theories that have been put forward to explain the origin of the obliquity\cite{alb22}, including:

\begin{figure}[tb]
\centering
\includegraphics[width=1.0\textwidth]{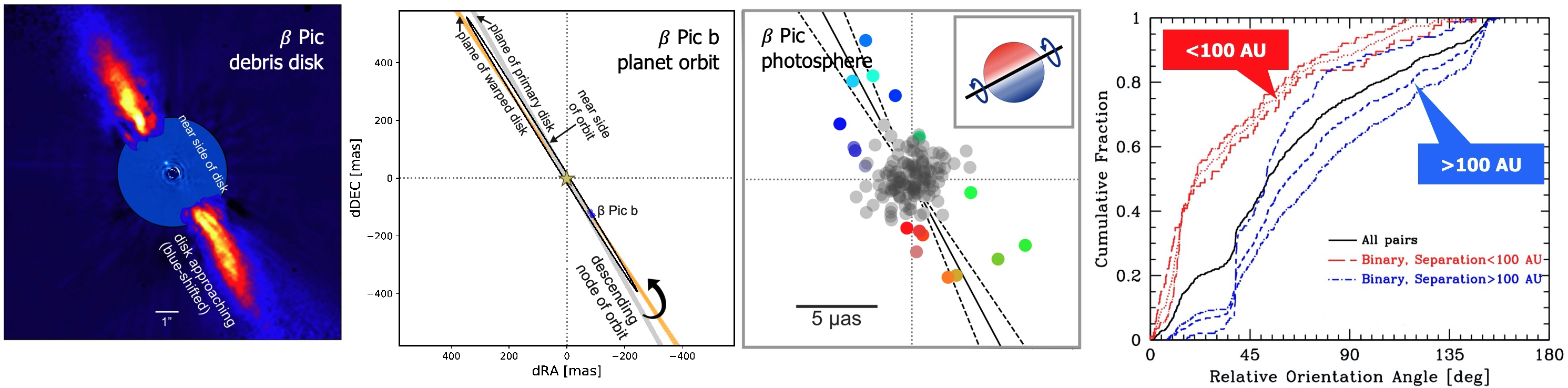} 
\caption{Left three panels: Spectro-interferometry in photospheric absorption lines constrains the sky-projected orientation of the stellar spin axis, as demonstrated for $\beta$\,Pic where the angular momentum vector of the star, the planet, and the debris disk were found to be aligned within 4$^{\circ}$ \cite{kra20a}. Right: Cloud-collapse simulations predict that the spins of close binaries should be predominantly aligned, while the spins of wide-separation systems should be more randomly distributed, probing the formation mechanism of these systems \cite{bat18}.}
\label{fig:spinorbit}
\end{figure}

\begin{itemize}
\item {\bf Multiple modes of formation:} Binaries formed through disk fragmentation should have their stellar spins and orbit spins predominantly aligned. For binaries formed through turbulent cloud fragmentation, the spin orientations should be distributed randomly. $\Rightarrow$ We will search for this predicted bimodality \cite{bat18} by comparing the spin-spin distributions for close ($<100$\,au) and wide ($>100$\,au) binaries. 

\item {\bf Kozai-Lidov mechanism:} A wide companion orbiting a close binary on a highly inclined orbit can induce oscillations in inclination/eccentricity of the close pair, possibly followed by orbital decay due to tidal friction \cite{fab07} $\Rightarrow$ We will be able to test this scenario by comparing the obliquity distribution of systems with an astrometrically-detected close-in companion/planet (i.e. separations $\lesssim$ few au) and a wide-separation companion (at a few hundred au) with the distribution of systems without wide-separation companions. 

\item {\bf Flyby of neighboring stars}\cite{gua04} We will test this hypothesis by measuring the obliquity distribution in clusters (e.g. Upper Sco), where stars close to the cluster core are more likely to have interacted with neighboring stars during their history than stars located in the periphery. Comparing the obliquity distribution for stars in the cluster core, in the cluster periphery, and of field stars, will inform us about the role of stellar encounters in inducing the obliquities.
\end{itemize}

We will use BIFROST’s R=25,000 mode in photospheric absorption lines to determine spin-orbit alignments. The sky-projected spin axis orientation can be constrained from the differential phase, providing the equivalent information as the RML effect. This method has been demonstrated on Fomalhaut \cite{leb09} and $\beta$\,Pic (\cite{kra20a}; Fig.~\ref{fig:spinorbit}). For stars that are seen near-pole on ($i>70^{\circ}$) the technique provides a non-detection – however, statistically less than 15\% of stars are seen at such high inclination. For individual systems we can also constrain the inclination based on $v \cdot \sin i$ and astroseismology \cite{lun14}. For systems without inclination constraints we will use Bayesian analysis methods that have been developed to interpret RML measurements to recover the underlying 3D obliquity distribution \cite{cam16}.
Depending on the separation, flux ratio, and diameters of the stars in a binary system, we will be able to detect the spin signatures of only the primary or the primary+secondary. In case the secondary is too distant/faint/small, we can measure the spin orientation of the primary and compare it to the orbit orientation ($\Rightarrow$ spin-orbit orientation). If the secondary is sufficiently close/bright/large, its spin orientation can be measured as well, either by separating the two phase signals spectrally using BIFROST’s R=25,000 mode, or by acquiring the secondary separately ($\Rightarrow$ spin-spin orientation).
Stellar spins are considered a good proxy for the angular momentum at the time of star formation and are unlikely to be affected by subsequent orbital evolution – therefore, spin-spin measurements with BIFROST offer a unique diagnostic to differentiate between competing models of binary formation (cloud fragmentation versus disk fragmentation). Spin-orbit alignments, on the other hand, trace the dynamical processes that shape the system architecture after formation (Kozai-Lidov, stellar flybys, body-body scattering, etc.). 

{\bf Importance to the field:} BIFROST is able to measure, for the first time, spin-spin and spin-orbit alignments for wide non-eclipsing systems, i.e. with separations from $\sim0.2$ to thousands of au. Such systems are completely inaccessible with RML measurements. Furthermore, BIFROST spin-orbit alignments can be obtained over the whole mass range from stellar-mass companions down to planets. This will reveal what mechanisms are common between binary formation and planet formation, and how they differ.

{\bf Advancements offered:} Gravity’s velocity resolution ($\Delta v \lesssim 65$\,km/s) is just marginally sufficient to resolve the pressure-broadened Br$\gamma$ line on fast rotating stars. BIFROST’s R=25,000 mode ($\Delta v=12$\,km/s) allows measuring (a) the stellar spin orientation in slower-rotating stars, and (b) the spin signatures in narrow atomic lines and for individual components in spectroscopic binaries.

\subsection{Mass accretion \& ejection from YSOs to AGN}
\label{sec:scienceaccretion}

\begin{figure}[b]
\centering
\includegraphics[width=1.0\textwidth]{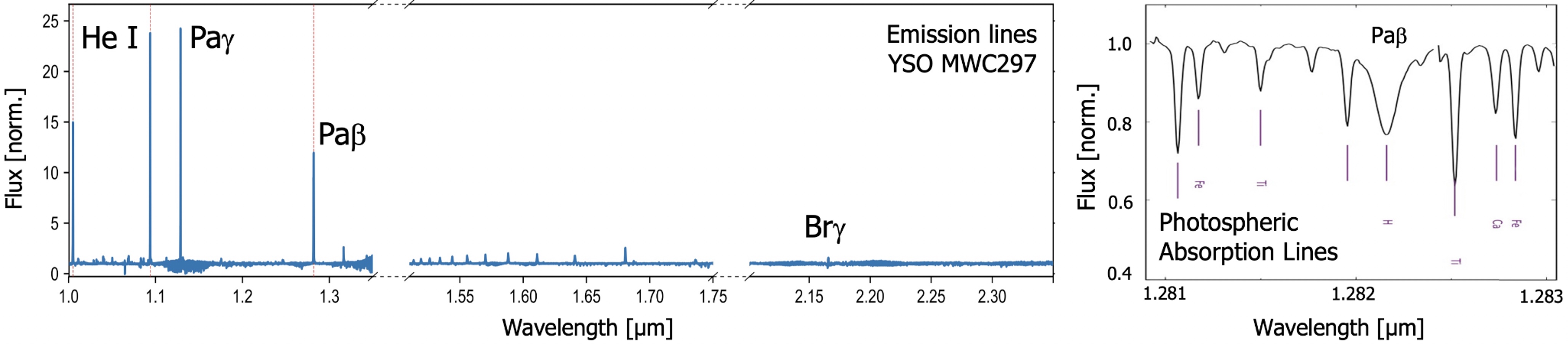} 
\caption{Left: The Y/J-band includes several unique line tracers in young stars, including HeI, Pa$\beta$\, and Pa$\gamma$. These lines have also a much higher equivalent width then the Br$\gamma$ line that is accessible in K-band (X-shooter spectrum of MWC297). Right: For the stellar spin measurements (science case 3.2), BIFROST’s R=25,000 mode allows us to measure the spin orientation in the pressure-broadened Pa$\beta$ line as well as in atomic lines. Fitting the rotation signatures in all lines simultaneously increases the SNR, allowing us to probe smaller stars (spectrum of 10 Leo; credit: Nicholls et al.\cite{nic17}).}
\label{fig:lines}
\end{figure}

\begin{figure}[t]
\centering
\includegraphics[width=1.0\textwidth]{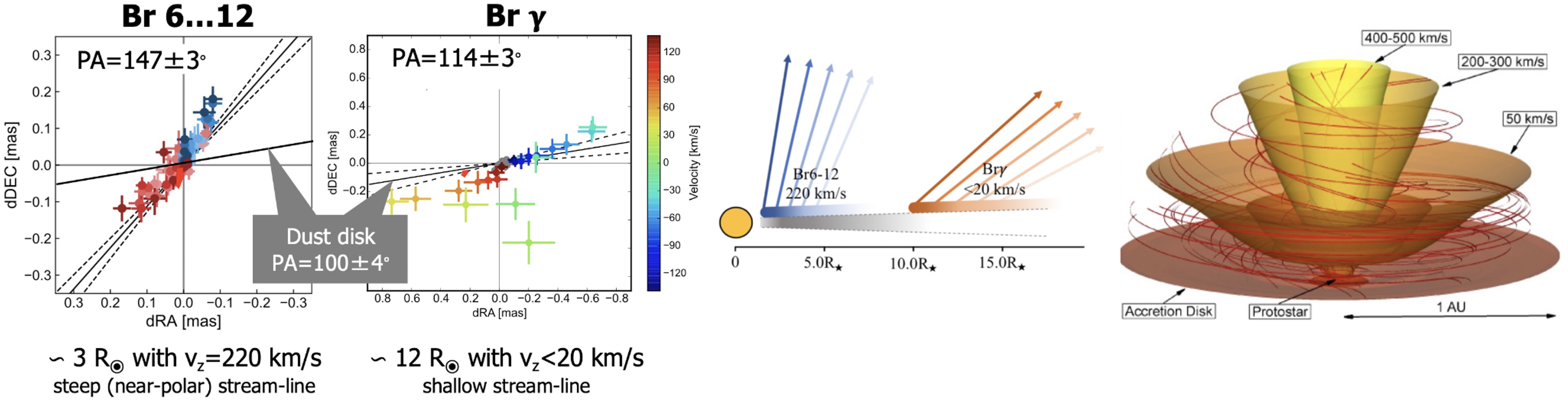} 
\caption{Left: Photocenter vectors measured on the Herbig B[e] star MWC297 with AMBER in the H-band (Br6…12) and K-band (Br$\gamma$ transition \cite{hon17}) reveal that the two line transitions originate from different spatial regions and follow substantially different sky-project velocity vectors (\cite{hon20}). Middle: Sketch of the 3D velocity vectors derived for these line transitions, suggesting that the wind collimation increases closer to the source. Right: Simulation of the magnetic-field geometry in a disk wind (credit: Romanova et al.\cite{rom15}), predicting the observed conical wind collimation effect. BIFROST will enable such gas kinematics measurements in better line tracers, at much higher spectral resolution, and for large object samples. }
\label{fig:accretion}
\end{figure}

Some of the VLTI’s most remarkable achievements have centered around measuring the kinematics in emission lines associated with a broad range of astrophysical objects, including accretion in T Tauri stars \cite{gra20}, disk winds around Herbig Ae/Be stars \cite{hon17}, wind-wind collision in luminous-blue variables \cite{wei16}, to relativistic jets in microquasars \cite{gra17} and Active Galactic Nuclei (AGN \cite{gra18}). However, the vast majority of these studies have been conducted in the K-band and the Br$\gamma$ hydrogen recombination line. There is much untapped potential in applying this technique to the Y/J/H-bands, which features a richer spectrum (Pa$\gamma$ 1.094\,$\mu$m, Pa$\beta$\,$1.282 \mu$m, Br6…Br12 1.73…1.55\,$\mu$m, [Fe II] 1.257\,$\mu$m, He\,I 1.083\,$\mu$m), and where the Paschen lines have 4...10-times higher equivalent width then Br$\gamma$ for typical young stars and AGN (Fig.~\ref{fig:lines}; see also Albrecht et al.\cite{alc14}).
To study circumstellar accretion, the He I 1.083\,$\mu$m line \cite{fis08} provides a much more reliable accretion tracer than Br$\gamma$. For young stars, it was found that Br$\gamma$ can be associated with magnetospheric accretion, disk winds, or both processes at the same time \cite{kra08b}. This often introduced uncertainties in the interpretation of Br$\gamma$ spectro-interferometric studies.
BIFROST’s R=25,000 mode in the He I line will provide a very direct accretion tracer, enabling studies of the accretion geometry near the Alven radius around nearby T Tauri and Herbig Ae/Be stars. Observations on PMS binaries will teach us how binaries accrete. For instance, it has been proposed that preferential accretion onto the secondary could drive the mass-ratio in binary systems towards unity and might explain the excess fraction of twin binaries that has been observed for close ($<0.1$\,au) binaries \cite{moe17}. We will be able to test this prediction by measuring individual accretion rates in PMS binaries.
By spatially \& spectrally resolving multiple line transitions of hydrogen recombination lines, BIFROST can constrain the physical conditions (gas density, temperature, excitation) and velocity field at different regions in the circumstellar environment. This will enable us to apply new approaches to reconstruct the 3-dimensional velocity field in the inner regions of protoplanetary disks, where accretion occurs and winds are launched from the star, the disk, or the interaction region between the stellar and disk magnetic field. Using differential phase information it is possible to separate the out-of-plane (poloidal) gas velocity component from the Keplerian (toroidal) velocity \cite{hon17}. Resolving multiple hydrogen line transitions (Pa$\beta$, Pa$\gamma$, Br6…Br12, and Br$\gamma$) will allow us to reconstruct the velocity structure at different radii in the disk, providing new diagnostics to constrain the magnetic field structure in the launching region of disk winds (Fig.~\ref{fig:accretion}).
BIFROST in J opens new opportunities for studying AGN in both Pa$\beta$ ($z<0.04$) and in H$\alpha$ ($0.7<z<1.0$), with $\sim 10$ Type 1 sources observable in each redshift range.  At low redshifts, the Paschen lines are in the J/H band, which limits Gravity+ studies to the relatively weak Br$\gamma$ line, where the constraints on kinematics of the Broad Line Region (BLR) are rather weak (differential phases $\lesssim 0.2^{\circ}$; Fig.~3 in Gravity+ collaboration\cite{grap}). For these AGN, BIFROST will achieve 2-times higher angular resolution in the stronger Pa$\beta$ line, resulting in higher-precision Black Hole masses and geometric scales for a range of cosmological studies. At redshifts approaching 1, the benefit of observing the redshifted strong H$\alpha$ line is even more pronounced. Spatial mapping of these emission lines will solve degeneracies on the geometry of the emitting-regions and the orientation/inclination of the BLR, allowing models of the BLR geometry to be calibrated before they are applied to samples at high redshift, where less information is available. Another Gravity+ science case is the detection of close Binary Supermassive Black Holes to solve the ‘final parsec problem’ in galaxy evolution. Their detection relies heavily on measuring and interpreting complex differential phase signatures \cite{son19}, which, for nearby AGN, can be achieved more robustly with BIFROST’s higher spectral resolution modes.\newline

{\bf Importance to the field:} Our immediate objectives are focussed on young stars and AGNs, but we note that this approach will also be applicable to study convection in stellar atmospheres, the decretion disks of classical Be stars, the mass-loss in AGB stars, Wolf-Rayet stars, or Luminous Blue Variables, or the jets of microquasars.\newline

{\bf Advancements offered:} BIFROST offers up to $6\times$ higher spectral resolution than Gravity and opens a spectral window of particular interest for line studies.

\subsection{Exoplanet Atmospheres and Circumplanetary Disks}
\label{sec:exoplanets}

Asgard’s off-axis fringe tracking mode will enable BIFROST to integrate up to 1.5” (for UTs; 4” for ATs) off-axis, while HEIMDALLR tracks on-axis on the star. This will allow us to place the BIFROST fiber at the position of an off-axis companion/planet, to adjust the internal differential delay lines, and to record deep interferograms at the predicted off-axis delay position. With the ATs it will enable spin alignment measurements of wide GAIA binaries, even if the companions are too faint for fringe tracking (science case 3.2). With the UTs it will allow us to characterize protoplanets and exoplanets whose position is known from direct imaging or inferred from GAIA DR3 astrometry. Gravity has used this method with much success for atmospheric characterisation of more than a dozen exoplanets in K-band, including HR8799e \cite{gra19}, $\beta$\,Pic b+c, and PDS70b. This approach combines the power of AO and interferometry, with a potential for achieving very high contrast near the diffraction limit of the UTs ($\sim0.025$''), enabling the following objectives:
\begin{itemize}
\item {\bf Gas kinematics in circumplanetary disks around protoplanets:} While a circumplanetary disk has now been detected around PDS70c with dust continuum imaging \cite{ben21} and H$\alpha$ spectroscopy \cite{haf19}, there is still substantial uncertainty on how protoplanets accrete. We will use BIFROST’s off-axis mode to search for differential phase signatures in the Pa$\gamma$ and Pa$\beta$ line for PDS70c-like protoplanets. Paschen lines are likely the sweet-spot for such studies, as simulations predict 5…10-times higher equivalent width for Pa$\beta$ than Br$\gamma$ \cite{aoy20,szu20} (possibly explaining the Br$\gamma$ non-detection reported with Gravity in Wang et al.\cite{wan21}), while the H$\alpha$ line is extremely susceptible to dust extinction and also unsuitable for interferometric follow-up with the existing VLTI infrastructure. Differential phase signatures measured with BIFROST could yield the first direct measurement of the mass of a protoplanet and provide insights on whether planets accrete through spherical infall, polar infall, or magnetospheric accretion \cite{mar22}. 

In order to evaluate the feasibility of observing circumplanetary disks with BIFROST, we estimated the Pa$\beta$ line flux of PDS70b to $L_{\textrm{Pa}\beta}/L_{\odot}=2.7\times10^{-8}$ based on the measured H$\alpha$ luminosity\cite{haf19} and planetary line-emission accretion models\cite{aoy21}. Assuming a similar line width as H$\alpha$ (line FWHM 100\,km/s), BIFROST's R=6000 off-axis mode should be able to obtain a 3$\sigma$ detection of the gas kinematic signatures of the circumplanetary disk around PDS70b as wavelength-differential phase signal within 5.9\,hrs observing time on the UTs.

\item {\bf Atmospheric retrieval for Jovian exoplanets:} Molecular tracers in Y/J/H-band probe deeper layers of the atmosphere than Gravity’s K-band and are essential to constrain cloud particle sizes\cite{que00} and the surface gravity. BIFROST’s R=1000+5000 mode  will be ideal to cover the broad O2 absorption bands in the J/H-band that are important for atmospheric retrieval. We expect that several dozen RV-detected giant planets and GAIA planets will be accessible. The combination of dynamical masses provided by these techniques and the spectra provided by BIFROST will be key to break atmosphere model degeneracies.
\end{itemize}

{\bf Importance to the field:} Enabling Y/J/H-band off-axis interferometry is urgent \& timely, as GAIA provides now a large sample of exoplanets to follow up on. Atmospheric retrieval studies with Gravity+’s K-band off-axis mode will face ambiguities unless complemented with short-wavelength spectroscopy (e.g. see Table 3 in Nowak et al.\cite{now20}). Also, JWST, ALMA, and ELT/METIS will likely discover new PDS70-like protoplanets around young stars, whose circumplanetary disks can be characterized with BIFROST. Resolving circumplanetary disks has been identified as a key science mission for a next-generation infrared interferometric facility (PFI concept study \cite{mon18}) and Asgard might be able to obtain first results for wide-separation protoplanets.\newline

{\bf Advancements offered and long-term perspective:} The off-axis mode will open immediate science applications in exoplanet spectroscopy and the measurement of microarcsecond-scale photocenter shifts to constrain the kinematics of gas in circumplanetary disks. 
Ambitious long-term developments could push in the following directions:
\begin{itemize}
\item The contrast at the inner working-angle will be determined largely by the Strehl of GPAO and of the Baldr AO. With highly optimized systems, e.g. a SCAR coronagraph \cite{por20} and multi-night integrations, it could be possible to achieve $10^{-5…-6}$ contrast needed for spectroscopy of reflected light on warm rocky planets.
\item  Our immediate goal is to enable off-axis correlated flux measurements (spectroscopy) and wavelength-differential astrometry in lines, which can be implemented as an Asgard-internal mode. However, there is potential for implementing a high-precision narrow angle (HPNA) astrometry mode for precision differential astrometry with respect to the fringe tracker phase reference, which would enable precision astrometry on exoplanets and the astrometric search for exomoons. This would require laser metrology out to the AT/UT mirrors and therefore necessitate deep involvement of ESO.
\end{itemize}

\section{INTEGRATION IN ASGARD SUITE}
\label{sec:asgard}

BIFROST is being proposed to ESO as part of a Suite of Instruments, namely the Asgard Suite of VLTI visitor instruments.
This Suite includes an image-plane fringe tracker, HEIMDALLR \cite{heimdallr_ireland}, and an adaptive optics system, Baldr.
The overall layout of the Asgard table is discussed in Martinod et al.\cite{asgard_martinod}, where the K-band light is used by HEIMDALLR 
for high-sensitivity, low-latency fringe tracking.  
Wavelength shorter than 2\,$\mu$m are reflected with a dichroic to Baldr and BIFROST.
A  second dichroic then splits either Y/J-band or H-band to Baldr, where the light is used to control deformable mirrors
early in the common Asgard beampath.  The other band, either H-band or Y/J-band, is transmitted to BIFROST.

\section{BIFROST OPTICAL DESIGN \& SUBSYSTEMS}
\label{sec:design}

\subsection{Two-arm design}
\label{sec:twoarms}

Our design choices are driven by the following objectives: (1) maximise the fringe contrast for YJ-band at high spectral resolution, (2) optimise the sensitivity for YJ-band observations by minimising the thermal background and by enabling long integration times in these band, and (3) at the same time enable spectro-interferometric observations in H-band at high spectral resolution.

This led us to adopt a two arm design, where the first arm (``YJH arm'') has a camera that is sensitive between 1 and 1.7\,$\mu$m.  For observations in H-band, a low-resolution prism, or medium/high-resolution grating is moved into the path to record wavelength-differential visibilities and phases on the YJH arm. For observations in YJ-band, the YJH arm is used with a low-dispersion prism (spectral resolution $R\approx50$) to monitor residual fringe drifts (e.g.\ due to chromatic dispersion caused by the moving delay lines), to absolute-calibrate the visibilities, and to enable frame selection in post-processing.  

For objects that are sufficiently bright, a beam splitter is moved into the beam and redirects the dominant fraction of the light (90\%) to a second arm (``YJ arm'') that hosts the high-spectral resolution gratings for Y and J-band and a camera with cut-off wavelength $\sim 1.4\,\mu$m.  The lower cut-off wavelength helps to reduce the thermal background and to achieve our sensitivity goals.  We plan for three grisms (R=1000, R=5000, R=25,000, corresponding to 300\,km/s, 50\,km/s, and 12\,km/s velocity resolution).  Using actuators it is possible to select wavelength settings either in the Y-band (He\,I 1.083\,$\mu$m, Pa$\gamma$ 1.094\,$\mu$m) or J-band ([Fe\,II] 1.257\,$\mu$m, Pa$\beta$ 1.282\,$\mu$m). By chosing an eAPD detector (Sect.~\ref{sec:detectors}), we can read out data from the YJ~arm with frame rates of several Hz, without significant penalities in terms of read-noise.  This allows us to use the group delay and flux information recorded on the YJH~arm for frame selection and post-processing, where we apply a phasor to the high-spectral resolution interferograms to correct for residual OPD drifts or fringe jumps between the exposures.  This should allow us to achieve high fringe contrast even for long effective integration times from minutes to an hour.

\begin{figure}[tb]
\centering
\includegraphics[width=1.0\textwidth]{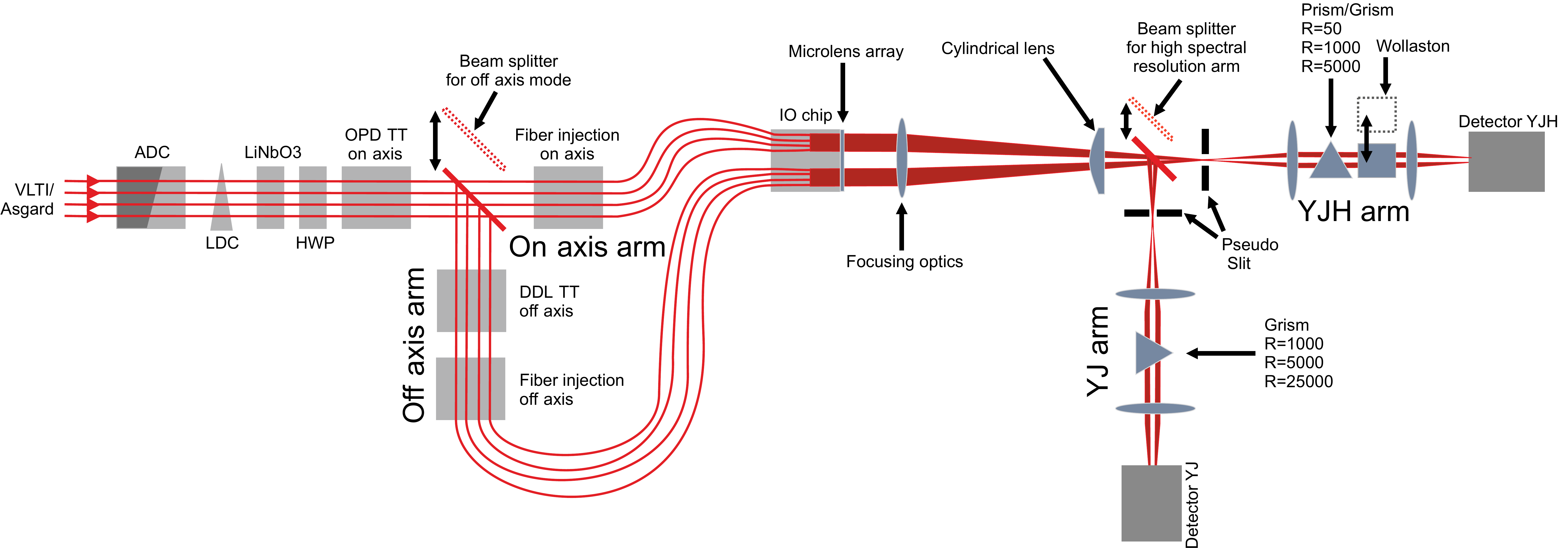} 
\caption{ Schematics of the beam path and components in BIFROST. }
\label{fig:design}
\end{figure}

\subsection{Ultra-low noise eAPD detectors}
\label{sec:detectors}

To reach our sensitivity goals we will make use of the recent breakthroughs in the field of electron avalanche photodiode (eAPD) detectors. 
The eAPD technology offers a truly revolutionary performance compared to the earlier-generation 
HAWAII or PICNIC detector arrays by incorporating an electron avalance multiplication stage that 
amplifies the signal before the image is read out.  So far, the read noise has been the dominant 
noise contributor for high-speed infrared detector required in astronomical interferometry.  
We plan to use Leonardo’s SAPHIRA chip\cite{fin16} that achieves a sub-electron read-noise at 3500\,Hz, 
as implemented in the CRED\,One camera from the company First Light Imaging\cite{fea22}.

In the medium-term, we plan to upgrade the detector on the YJ arm with a large-format eAPD camera (2048$\times$2048 pixels). 
That camera might also include a cold filterwheel, which would allow one to choose for each spectral setting a cold filter with 
appropriate cut-off wavelength. In this case, we could then accommodate all high spectral resolution gratings for Y, J, and 
H-band on the second arm, leaving the first arm only with the low-spectral dispersion prism.  
This upgrade would increasing the wavelength range and  reduce the number of
operating modes. 

\subsection{Dual-field mode}
\label{sec:dualfield}

The objective of the off-axis mode is to enable spectroscopy (correlated flux measurements) of faint companions and to characterise the environment around companions with wavelength-differential visibilities and phases.  As mentioned before, the strength of this method is that it combines two powerful methods for suppressing starlight, namely the reduction of photon noise through extreme adaptive optics (to be implemented on all UTs as part of the ongoing Gravity+ infrastructure improvements) and filtering of residual starlight based on the interferometry phase information\cite{now20}.

To enable this mode, BIFROST will have two set of optical fibres that can be placed at arbitrary positions within the field-of-view of the ATs or UTs.  Typically, one fiber will be placed on the star on-axis, while the second fibre will be placed on the expected position of the much fainter off-axis companion.  We install differential delay lines in front of the off-axis fiber injection module. The DLL position is adjusted to account for the phase difference that corresponds to the angular positions on the sky, allowing us to carry out deep blind integrations at the expected OPD position of the fringe package of the off-axis source.  

A key requirement for this mode is to measure the phase difference between the on-axis and off-axis source with high precision, which requires on the one hand to measure the phase of the on-axis and off-axis fringe simultaneously, and on the other hand, to control the optical path difference between the place where these phases are measured, to sub-wavelength precision.  In the Gravity beam combiner, the phases of the on-axis source and the off-axis source are measured in two separate instrument modules, namely the fringe tracker (Gravity-FT) and the science combiner (Gravity-SC).  A laser metrology system is employed to monitor the precise OPD between these two combiners, so that correlated flux and precision astrometry of the off-axis source can be derived down to scales of a few tens of microarcseconds.

For BIFROST, we try to minimise any potential OPD drifts between the on-axis and off-axis source from the start, namely by placing the interferograms of on-axis and off-axis source side-by-side on the same detector.  The light passes through the same integrated optics combiner and the same spectrograph, allowing the phases of the on-axis and off-axis source to be measured perfectly simultaneously.  There are only very few optical components that differ on the two beampaths, namely the two fiber-injection modules and the differential delay line.  Chosing high-precision translation stages for the differential delay line should therefore allow us to control the OPD between the on-axis and off-axis source to sub-wavelength precision, meeting the requirement to conduct exoplanet spectroscopy and potentially also astrometry.  We are still analysing potential systematic error source and will consider adding a metrology system to improve the accuracy further.

\subsection{Subsystems}
\label{sec:subsystems}

The optical layout of BIFROST is illustrated in Fig.~\ref{fig:design} and contains the following components:

\begin{itemize}
\item {\bf Atmospheric Dispersion Corrector (ADC):}  Ensures that light over the full wavelength bandpass (either Y/J-band or H-band) can be coupled into the fiber.   As shown in Mortimer et al.\cite{bifrost_mortimer}, atmospheric dispersion could constitute a significant problem for observations with the UTs at high airmass, where the angular displacement could reach up to $\sim 150$\,mas between 1.0 and 1.4\,$\mu$m, far larger than the diffraction-limited resolution of the UTs ($\sim 30$\,mas).

\item {\bf Longitudinal Dispersion Corrector (LDC):} Ensures that the optical path delay is matched across the full bandwidth. 

\item {\bf Birefringence Correction (LiNbO3 plates):} Corrects for differential birefringence between the different lightpaths to enable a polarisation-insensitive measurement of the fringe contrast.  Following Lazareff et al.\cite{laz12}, we plan to implement this with rotatable Lithium Niobate (LiNbO3) plates, whose rotation is adjusted to balance the birefringence between the beams.

\item {\bf Half-wave plates (HWP plates):} We will account for space to install half-wave plates in the beam that can be rotated to measure different polarisation states.  This could enable polarisation-sensitive interferometric observations, e.g.\ to characterise the dust properties in the inner regions of protoplanetary disks, although it will require additional efforts to characterise and calibrate the polarisation properties of the VLTI infrastructure.

\item {\bf OPD TT correction on-axis:} This unit will likely be implemented through mirrors on a translation stage and a fast stearing mirror and will allow to adjust the optical path delay (OPD) for each beam, as well as the light injection of the on-axis starlight into the fibers.

\item {\bf Beam splitter off-axis mode:} This beam splitter can be inserted into the beam to split off light for off-axis observations. Given that the off-axis source is typically much fainter than the on-axis starlight, we consider to use a highly asymmetric beam splitter that transmits only a few percent of the on-axis star light  -- just enough so that the relative phase between the star and the off-axis source can be monitored.

\item {\bf Fiber-injection module on-axis:} An off-axis parabloid (OAP) to inject the on-axis light into fibers.  We might have separate fibers for YJ-band and H-band.  The fiber mounts will be positioned on a motorised stage, so that either the set of YJ-band fibers or the set of H-band fibers can be moved into the focus of the OAPs.

\item {\bf DDL TT correction off-axis:} This unit will have a similar design as the ``OPD TT correction on-axis'' unit and act as differential delay line (DDL) between the on-axis and off-axis source.  The actuators will be chosen for optimal precision and reproducibility, so that the differential delay for the off-axis recording can be adjusted with sub-wavelength precision.

\item {\bf Fiber-injection module off-axis:} This unit will have a similar design as the ``Fiber-injection module on-axis'' unit and inject the light at the off-axis position into single-mode fibers.

\item {\bf Integrated optics combiner (IO device):} An integrated optics circuit that will combine the light pairwise, using the ABCD method to sample the fringe signal on the six baselines.   We will acquire a custom-made integrated optics device, where the 24 outputs from the on-axis source and the 24 outputs from the off-axis source are positioned next to each other, so that they can be recorded simultaneously.  A microlens array that is clued onto the device collimates the light from the IO outputs.  Given that IO combiners can be optimized only for a limited bandwidth, we plan for two IO devices -- one for the YJ-band and a separate one for H-band.  The two devices are mounted on a motorized stage, so that either the YJ-band or H-band device can feed light into the spectrograph.  Details on the design and an alternative All-In-One combiner design are described in Mortimer et al.\cite{bifrost_mortimer}.

\item {\bf Spectrograph:} The spectrograph consists of focusing optics, a cylindrial lens that compresses the beam so that it can be spectrally dispersed.  The dispersing elements on the YJH arm are a low spectral resolution prism ($R\approx50$, although the precise value will still be chosen based on interferometric field-of-view and sensitivity considerations) and gratings for the H-band (R=1000 and R=5000).  On the YJ arm, we plan for 3 gratings (R=1000, R=5000, R=25,000), where the considered grating technologies (volume Phase Holographic gratings and Binary gratings) are discussed in \cite{bifrost_chhabra}.

\item {\bf Beam Splitter for high-resolution YJ arm:} For high-spectral resolution observations, an asymmetric beam splitter can be moved into the beampath to redirect $\sim 90$\% of the light to the YJ arm.

\item {\bf Detectors:} As outlined in section~\ref{sec:detectors}, we plan to use eAPD detectors optimized for low read-noise and low dark current on both arms.
\end{itemize}

\begin{table}[b]
\caption{BIFROST operating modes}
\label{tab:modes}
\centering
\vspace{0.1cm}
\begin{tabular}{c c c c c c c}
\hline\hline
Instrument   &  Mode           & Spectral    & Spectral   & Spectral   & Beam & Comment \\
Arm              &                       & Band        & resolution & range$^{a}$ [$\mu$m] & splitter$^{b}$ &  \\
 \hline
Arm~YJH           &  LR-YJ  &  YJ     &50  & & IN or OUT &  \\
Arm~YJH           &  LR-H  &  H     & 50  & & OUT &  \\
Arm~YJH          &  LR-YJ-WOLL    &  YJ          & 50  & & IN or OUT & Split polarisation\\
Arm~YJH          &  LR-H-WOLL    &  H          & 50  & & OUT & Split polarisation\\
Arm~YJH         &  MR-H           &  H        & 1000 & 1.28-1.77 & OUT &  \\
Arm~YJH         &  HR-H           &  H        & 5000 & 1.56-1.66 & OUT &  \\
\hline
\multicolumn{7}{c}{To record YJ~band high-spectral resolution data on Arm~YJ, the beam splitter needs to be IN}\\
\multicolumn{7}{c}{and Arm~YJH needs to be in LR-YJ or LR-YJ-WOLL setup.} \\
\hline
Arm~YJ           &  MR-YJ          &  YJ        & 1000 & 1.00-1.38 & IN & \\
Arm~YJ          &  HR-Y            &  Y         & 5000 & 1.06-1.13 & IN & \\
Arm~YJ         &  HR -J            &  J           & 5000 & 1.22-1.30 & IN & \\
Arm~YJ         &  VHR-PaG-HeI       & Y              & 25,000 & 1.082-1.095 & IN &  He\,I 1.083 \,$\mu$m, \\
              &             &                &  &  & &  Pa$\gamma$ 1.094\,$\mu$m\\
Arm~YJ         &  VHR-[FeII]             & J              & 25,000 & 1.249-1.265  & IN & [Fe II] 1.257\,$\mu$m\\
Arm~YJ         &  VHR-PaB                & J              & 25,000 & 1.273-1.290  & IN & Pa$\beta$ 1.282\,$\mu$m\\
 \hline
\end{tabular}\\

\footnotesize{$^a$ The spectral range has been computed for a 320$\times$256 SAPHIRA detector, but could be increased up to 6-fold with existing 2048$\times$2048 eAPD technology. } \footnotesize{$^b$ The spectrograph beam splitter is OUT for the HIGHSENS mode and for observations in H-band.}\\
\end{table}

\section{PLANNED OPERATING MODES}
\label{sec:modes}

To match the requirements posed by different science target stars,
BIFROST will offer different operating modes that allow one to select different 
spectral bands (Y-band, J-band, YJ-band, or H-band),
windows around specific spectral lines, or to optimise the
instrument for high sensitivity at low spectral resolution (``HIGHSENS'' mode).
Also, we plan for a split-polarisation mode that can be used to 
measure the two polarisation states separately (``WOLL'' mode).
This mode could be used for improving the visibility calibration and, potentially,
also for future science observations in polarized light, although this will require
detailed modelling of the polisation properties long the VLTI beam path.

These different set ups are achieved by moving the spectrograph beam-splitter
into the beam (``IN'' state, for redirecting 90\% of the light to the YJ~arm for high-spectral
resolution observations in Y and/or J-band) or out of the beam (for HIGHSENS observations
or observations in H-band).  The filterwheels be used to move different 
dispersion elements into the beampaths, where the YJH arm host medium/high-resolution 
gratings for the H-band (MR-H, R=1000; HR-H, R=5000) and the low-dispersion
prisms that are needed either for HIGHSENS observations to monitor flux dropouts, 
chromatic dispersion, and fringe jumps (see Sect.~\ref{sec:twoarms}) for simultaneous observations on the YJ~arm.
The filterwheel on the YJ~arm hosts medium (MR-YJ, R=1000), high (HR-Y, HR-J, R=5000), and 
very high-resolution (R=25,000) gratings for the YJ-band, where the actuators on the beam splitter 
can be used to adjust the spectral window on the detector.
For the WOLL mode, we move a Wollaston prism into the beampath on the YJH~arm.

We summarise the anticipated operating modes in Table~\ref{tab:modes}.

\section{CONCLUSIONS}
\label{sec:conclusions}

The BIFROST instrument described here will benefit applications in science areas that range from measuring the fundamental stellar parameters, architecture \& spin-orbit alignment of GAIA binaries to studying accretion/ejection in young stars, evolved stars, to Active Galactic Nuclei at high spectral resolution. Furthermore, BIFROST's off-axis mode will enable unique new science on characterising exoplanet atmospheres and measuring the gas kinematics in circumplanetary disks around young protoplanets.  

We work towards finalising the optical design in the first quarter of 2023 and will start the integration of the instrument immediately thereafter.  In March 2022, we initiated the approval process for bringing BIFROST and the other Asgard instruments to the VLTI visitor focus.  Further information and the latest news on our instrumentation efforts can be found at \url{http://bifrost.skraus.eu}.

\acknowledgments     

We are grateful for the constructive interactions with colleagues at ESO, in particular Fr\'{e}d\'{e}ric Gonte, Xavier Haubois, Antoine M\'{e}rand, Nicolas Schuhler, and Julien Woillez.  We  acknowledge support from an European Research Council (ERC) Consolidator Grant (``GAIA-BIFROST'', grant agreement No.\ 101003096) and STFC Consolidated Grant (ST/V000721/1).  
D.D. acknowledges support from the ERC under the European Union's Horizon 2020 research and innovation program (grant agreement No.\ CoG - 866070). M-A.M. has received funding from the European Union’s Horizon 2020 research and innovation programme under grant agreement No.\ 101004719.


\bibliography{BIFROST}   

\begin{thebibliography}{10}

\bibitem{fis08}
W.~{Fischer}, J.~{Kwan}, S.~{Edwards}, and L.~{Hillenbrand}, ``{Redshifted
  Absorption at He I {\ensuremath{\lambda}}10830 as a Probe of the Accretion
  Geometry of T Tauri Stars},'' {\em \apj}~{\bf 687}, pp.~1117--1144, Nov.
  2008.

\bibitem{kra19}
S.~{Kraus}, ``{Star Formation and Fundamental Stellar Astrophysics with VLTI at
  1 micron},'' in {\em The Very Large Telescope in 2030},  p.~36, July 2019.

\bibitem{cas08}
S.~{Casertano}, M.~G. {Lattanzi}, A.~{Sozzetti}, A.~{Spagna}, S.~{Jancart},
  R.~{Morbidelli}, R.~{Pannunzio}, D.~{Pourbaix}, and D.~{Queloz},
  ``{Double-blind test program for astrometric planet detection with Gaia},''
  {\em \aap}~{\bf 482}, pp.~699--729, May 2008.

\bibitem{gai16}
{Gaia Collaboration}, A.~G.~A. {Brown}, A.~{Vallenari}, T.~{Prusti}, J.~H.~J.
  {de Bruijne}, F.~{Mignard}, R.~{Drimmel}, C.~{Babusiaux}, C.~A.~L.
  {Bailer-Jones}, U.~{Bastian}, M.~{Biermann}, D.~W. {Evans}, L.~{Eyer},
  F.~{Jansen}, C.~{Jordi}, D.~{Katz}, S.~A. {Klioner}, U.~{Lammers},
  L.~{Lindegren}, X.~{Luri}, W.~{O'Mullane}, C.~{Panem}, D.~{Pourbaix},
  S.~{Randich}, P.~{Sartoretti}, H.~I. {Siddiqui}, C.~{Soubiran}, V.~{Valette},
  F.~{van Leeuwen}, N.~A. {Walton}, C.~{Aerts}, F.~{Arenou}, M.~{Cropper},
  E.~{H{\o}g}, M.~G. {Lattanzi}, E.~K. {Grebel}, A.~D. {Holland}, C.~{Huc},
  X.~{Passot}, M.~{Perryman}, L.~{Bramante}, C.~{Cacciari}, J.~{Casta{\~n}eda},
  L.~{Chaoul}, N.~{Cheek}, F.~{De Angeli}, C.~{Fabricius}, R.~{Guerra},
  J.~{Hern{\'a}ndez}, A.~{Jean-Antoine-Piccolo}, E.~{Masana}, R.~{Messineo},
  N.~{Mowlavi}, K.~{Nienartowicz}, D.~{Ord{\'o}{\~n}ez-Blanco}, P.~{Panuzzo},
  J.~{Portell}, P.~J. {Richards}, M.~{Riello}, G.~M. {Seabroke}, P.~{Tanga},
  F.~{Th{\'e}venin}, J.~{Torra}, S.~G. {Els}, G.~{Gracia-Abril},
  G.~{Comoretto}, M.~{Garcia-Reinaldos}, T.~{Lock}, E.~{Mercier}, M.~{Altmann},
  R.~{Andrae}, T.~L. {Astraatmadja}, I.~{Bellas-Velidis}, K.~{Benson},
  J.~{Berthier}, R.~{Blomme}, G.~{Busso}, B.~{Carry}, A.~{Cellino},
  G.~{Clementini}, S.~{Cowell}, O.~{Creevey}, J.~{Cuypers}, M.~{Davidson},
  J.~{De Ridder}, A.~{de Torres}, L.~{Delchambre}, A.~{Dell'Oro},
  C.~{Ducourant}, Y.~{Fr{\'e}mat}, M.~{Garc{\'\i}a-Torres}, E.~{Gosset}, J.~L.
  {Halbwachs}, N.~C. {Hambly}, D.~L. {Harrison}, M.~{Hauser}, D.~{Hestroffer},
  S.~T. {Hodgkin}, H.~E. {Huckle}, A.~{Hutton}, G.~{Jasniewicz}, S.~{Jordan},
  M.~{Kontizas}, A.~J. {Korn}, A.~C. {Lanzafame}, M.~{Manteiga}, A.~{Moitinho},
  K.~{Muinonen}, J.~{Osinde}, E.~{Pancino}, T.~{Pauwels}, J.~M. {Petit},
  A.~{Recio-Blanco}, A.~C. {Robin}, L.~M. {Sarro}, C.~{Siopis}, M.~{Smith},
  K.~W. {Smith}, A.~{Sozzetti}, W.~{Thuillot}, W.~{van Reeven}, Y.~{Viala},
  U.~{Abbas}, A.~{Abreu Aramburu}, S.~{Accart}, J.~J. {Aguado}, P.~M. {Allan},
  W.~{Allasia}, G.~{Altavilla}, M.~A. {{\'A}lvarez}, J.~{Alves}, R.~I.
  {Anderson}, A.~H. {Andrei}, E.~{Anglada Varela}, E.~{Antiche}, T.~{Antoja},
  S.~{Ant{\'o}n}, B.~{Arcay}, N.~{Bach}, S.~G. {Baker},
  L.~{Balaguer-N{\'u}{\~n}ez}, C.~{Barache}, C.~{Barata}, A.~{Barbier},
  F.~{Barblan}, D.~{Barrado y Navascu{\'e}s}, M.~{Barros}, M.~A. {Barstow},
  U.~{Becciani}, M.~{Bellazzini}, A.~{Bello Garc{\'\i}a}, V.~{Belokurov},
  P.~{Bendjoya}, A.~{Berihuete}, L.~{Bianchi}, O.~{Bienaym{\'e}},
  F.~{Billebaud}, N.~{Blagorodnova}, S.~{Blanco-Cuaresma}, T.~{Boch},
  A.~{Bombrun}, R.~{Borrachero}, S.~{Bouquillon}, G.~{Bourda}, H.~{Bouy},
  A.~{Bragaglia}, M.~A. {Breddels}, N.~{Brouillet}, T.~{Br{\"u}semeister},
  B.~{Bucciarelli}, P.~{Burgess}, R.~{Burgon}, A.~{Burlacu}, D.~{Busonero},
  R.~{Buzzi}, E.~{Caffau}, J.~{Cambras}, H.~{Campbell}, R.~{Cancelliere},
  T.~{Cantat-Gaudin}, T.~{Carlucci}, J.~M. {Carrasco}, M.~{Castellani},
  P.~{Charlot}, J.~{Charnas}, A.~{Chiavassa}, M.~{Clotet}, G.~{Cocozza}, R.~S.
  {Collins}, G.~{Costigan}, F.~{Crifo}, N.~J.~G. {Cross}, M.~{Crosta},
  C.~{Crowley}, C.~{Dafonte}, Y.~{Damerdji}, A.~{Dapergolas}, P.~{David},
  M.~{David}, P.~{De Cat}, F.~{de Felice}, P.~{de Laverny}, F.~{De Luise},
  R.~{De March}, D.~{de Martino}, R.~{de Souza}, J.~{Debosscher}, E.~{del
  Pozo}, M.~{Delbo}, A.~{Delgado}, H.~E. {Delgado}, P.~{Di Matteo},
  S.~{Diakite}, E.~{Distefano}, C.~{Dolding}, S.~{Dos Anjos}, P.~{Drazinos},
  J.~{Duran}, Y.~{Dzigan}, B.~{Edvardsson}, H.~{Enke}, N.~W. {Evans},
  G.~{Eynard Bontemps}, C.~{Fabre}, M.~{Fabrizio}, S.~{Faigler}, A.~J.
  {Falc{\~a}o}, M.~{Farr{\`a}s Casas}, L.~{Federici}, G.~{Fedorets},
  J.~{Fern{\'a}ndez-Hern{\'a}ndez}, P.~{Fernique}, A.~{Fienga}, F.~{Figueras},
  F.~{Filippi}, K.~{Findeisen}, A.~{Fonti}, M.~{Fouesneau}, E.~{Fraile},
  M.~{Fraser}, J.~{Fuchs}, M.~{Gai}, S.~{Galleti}, L.~{Galluccio},
  D.~{Garabato}, F.~{Garc{\'\i}a-Sedano}, A.~{Garofalo}, N.~{Garralda},
  P.~{Gavras}, J.~{Gerssen}, R.~{Geyer}, G.~{Gilmore}, S.~{Girona},
  G.~{Giuffrida}, M.~{Gomes}, A.~{Gonz{\'a}lez-Marcos},
  J.~{Gonz{\'a}lez-N{\'u}{\~n}ez}, J.~J. {Gonz{\'a}lez-Vidal}, M.~{Granvik},
  A.~{Guerrier}, P.~{Guillout}, J.~{Guiraud}, A.~{G{\'u}rpide},
  R.~{Guti{\'e}rrez-S{\'a}nchez}, L.~P. {Guy}, R.~{Haigron},
  D.~{Hatzidimitriou}, M.~{Haywood}, U.~{Heiter}, A.~{Helmi}, D.~{Hobbs},
  W.~{Hofmann}, B.~{Holl}, G.~{Holland}, J.~A.~S. {Hunt}, A.~{Hypki},
  V.~{Icardi}, M.~{Irwin}, G.~{Jevardat de Fombelle}, P.~{Jofr{\'e}}, P.~G.
  {Jonker}, A.~{Jorissen}, F.~{Julbe}, A.~{Karampelas}, A.~{Kochoska},
  R.~{Kohley}, K.~{Kolenberg}, E.~{Kontizas}, S.~E. {Koposov}, G.~{Kordopatis},
  P.~{Koubsky}, A.~{Krone-Martins}, M.~{Kudryashova}, I.~{Kull}, R.~K.
  {Bachchan}, F.~{Lacoste-Seris}, A.~F. {Lanza}, J.~B. {Lavigne}, C.~{Le
  Poncin-Lafitte}, Y.~{Lebreton}, T.~{Lebzelter}, S.~{Leccia}, N.~{Leclerc},
  I.~{Lecoeur-Taibi}, V.~{Lemaitre}, H.~{Lenhardt}, F.~{Leroux}, S.~{Liao},
  E.~{Licata}, H.~E.~P. {Lindstr{\o}m}, T.~A. {Lister}, E.~{Livanou},
  A.~{Lobel}, W.~{L{\"o}ffler}, M.~{L{\'o}pez}, D.~{Lorenz}, I.~{MacDonald},
  T.~{Magalh{\~a}es Fernandes}, S.~{Managau}, R.~G. {Mann}, G.~{Mantelet},
  O.~{Marchal}, J.~M. {Marchant}, M.~{Marconi}, S.~{Marinoni}, P.~M. {Marrese},
  G.~{Marschalk{\'o}}, D.~J. {Marshall}, J.~M. {Mart{\'\i}n-Fleitas},
  M.~{Martino}, N.~{Mary}, G.~{Matijevi{\v{c}}}, T.~{Mazeh}, P.~J. {McMillan},
  S.~{Messina}, D.~{Michalik}, N.~R. {Millar}, B.~M.~H. {Miranda}, D.~{Molina},
  R.~{Molinaro}, M.~{Molinaro}, L.~{Moln{\'a}r}, M.~{Moniez}, P.~{Montegriffo},
  R.~{Mor}, A.~{Mora}, R.~{Morbidelli}, T.~{Morel}, S.~{Morgenthaler},
  D.~{Morris}, A.~F. {Mulone}, T.~{Muraveva}, I.~{Musella}, J.~{Narbonne},
  G.~{Nelemans}, L.~{Nicastro}, L.~{Noval}, C.~{Ord{\'e}novic},
  J.~{Ordieres-Mer{\'e}}, P.~{Osborne}, C.~{Pagani}, I.~{Pagano}, F.~{Pailler},
  H.~{Palacin}, L.~{Palaversa}, P.~{Parsons}, M.~{Pecoraro}, R.~{Pedrosa},
  H.~{Pentik{\"a}inen}, B.~{Pichon}, A.~M. {Piersimoni}, F.~X. {Pineau},
  E.~{Plachy}, G.~{Plum}, E.~{Poujoulet}, A.~{Pr{\v{s}}a}, L.~{Pulone},
  S.~{Ragaini}, S.~{Rago}, N.~{Rambaux}, M.~{Ramos-Lerate}, P.~{Ranalli},
  G.~{Rauw}, A.~{Read}, S.~{Regibo}, C.~{Reyl{\'e}}, R.~A. {Ribeiro},
  L.~{Rimoldini}, V.~{Ripepi}, A.~{Riva}, G.~{Rixon}, M.~{Roelens},
  M.~{Romero-G{\'o}mez}, N.~{Rowell}, F.~{Royer}, L.~{Ruiz-Dern},
  G.~{Sadowski}, T.~{Sagrist{\`a} Sell{\'e}s}, J.~{Sahlmann}, J.~{Salgado},
  E.~{Salguero}, M.~{Sarasso}, H.~{Savietto}, M.~{Schultheis}, E.~{Sciacca},
  M.~{Segol}, J.~C. {Segovia}, D.~{Segransan}, I.~C. {Shih}, R.~{Smareglia},
  R.~L. {Smart}, E.~{Solano}, F.~{Solitro}, R.~{Sordo}, S.~{Soria Nieto},
  J.~{Souchay}, A.~{Spagna}, F.~{Spoto}, U.~{Stampa}, I.~A. {Steele},
  H.~{Steidelm{\"u}ller}, C.~A. {Stephenson}, H.~{Stoev}, F.~F. {Suess},
  M.~{S{\"u}veges}, J.~{Surdej}, L.~{Szabados}, E.~{Szegedi-Elek},
  D.~{Tapiador}, F.~{Taris}, G.~{Tauran}, M.~B. {Taylor}, R.~{Teixeira},
  D.~{Terrett}, B.~{Tingley}, S.~C. {Trager}, C.~{Turon}, A.~{Ulla},
  E.~{Utrilla}, G.~{Valentini}, A.~{van Elteren}, E.~{Van Hemelryck}, M.~{van
  Leeuwen}, M.~{Varadi}, A.~{Vecchiato}, J.~{Veljanoski}, T.~{Via},
  D.~{Vicente}, S.~{Vogt}, H.~{Voss}, V.~{Votruba}, S.~{Voutsinas},
  G.~{Walmsley}, M.~{Weiler}, K.~{Weingrill}, T.~{Wevers}, {\L}.~{Wyrzykowski},
  A.~{Yoldas}, M.~{{\v{Z}}erjal}, S.~{Zucker}, C.~{Zurbach}, T.~{Zwitter},
  A.~{Alecu}, M.~{Allen}, C.~{Allende Prieto}, A.~{Amorim},
  G.~{Anglada-Escud{\'e}}, V.~{Arsenijevic}, S.~{Azaz}, P.~{Balm}, M.~{Beck},
  H.~H. {Bernstein}, L.~{Bigot}, A.~{Bijaoui}, C.~{Blasco}, M.~{Bonfigli},
  G.~{Bono}, S.~{Boudreault}, A.~{Bressan}, S.~{Brown}, P.~M. {Brunet},
  P.~{Bunclark}, R.~{Buonanno}, A.~G. {Butkevich}, C.~{Carret}, C.~{Carrion},
  L.~{Chemin}, F.~{Ch{\'e}reau}, L.~{Corcione}, E.~{Darmigny}, K.~S. {de Boer},
  P.~{de Teodoro}, P.~T. {de Zeeuw}, C.~{Delle Luche}, C.~D. {Domingues},
  P.~{Dubath}, F.~{Fodor}, B.~{Fr{\'e}zouls}, A.~{Fries}, D.~{Fustes},
  D.~{Fyfe}, E.~{Gallardo}, J.~{Gallegos}, D.~{Gardiol}, M.~{Gebran},
  A.~{Gomboc}, A.~{G{\'o}mez}, E.~{Grux}, A.~{Gueguen}, A.~{Heyrovsky},
  J.~{Hoar}, G.~{Iannicola}, Y.~{Isasi Parache}, A.~M. {Janotto}, E.~{Joliet},
  A.~{Jonckheere}, R.~{Keil}, D.~W. {Kim}, P.~{Klagyivik}, J.~{Klar},
  J.~{Knude}, O.~{Kochukhov}, I.~{Kolka}, J.~{Kos}, A.~{Kutka}, V.~{Lainey},
  D.~{LeBouquin}, C.~{Liu}, D.~{Loreggia}, V.~V. {Makarov}, M.~G. {Marseille},
  C.~{Martayan}, O.~{Martinez-Rubi}, B.~{Massart}, F.~{Meynadier}, S.~{Mignot},
  U.~{Munari}, A.~T. {Nguyen}, T.~{Nordlander}, P.~{Ocvirk}, K.~S.
  {O'Flaherty}, A.~{Olias Sanz}, P.~{Ortiz}, J.~{Osorio}, D.~{Oszkiewicz},
  A.~{Ouzounis}, M.~{Palmer}, P.~{Park}, E.~{Pasquato}, C.~{Peltzer},
  J.~{Peralta}, F.~{P{\'e}turaud}, T.~{Pieniluoma}, E.~{Pigozzi}, J.~{Poels},
  G.~{Prat}, T.~{Prod'homme}, F.~{Raison}, J.~M. {Rebordao}, D.~{Risquez},
  B.~{Rocca-Volmerange}, S.~{Rosen}, M.~I. {Ruiz-Fuertes}, F.~{Russo},
  S.~{Sembay}, I.~{Serraller Vizcaino}, A.~{Short}, A.~{Siebert}, H.~{Silva},
  D.~{Sinachopoulos}, E.~{Slezak}, M.~{Soffel}, D.~{Sosnowska},
  V.~{Strai{\v{z}}ys}, M.~{ter Linden}, D.~{Terrell}, S.~{Theil}, C.~{Tiede},
  L.~{Troisi}, P.~{Tsalmantza}, D.~{Tur}, M.~{Vaccari}, F.~{Vachier},
  P.~{Valles}, W.~{Van Hamme}, L.~{Veltz}, J.~{Virtanen}, J.~M. {Wallut},
  R.~{Wichmann}, M.~I. {Wilkinson}, H.~{Ziaeepour}, and S.~{Zschocke}, ``{Gaia
  Data Release 1. Summary of the astrometric, photometric, and survey
  properties},'' {\em \aap}~{\bf 595}, p.~A2, Nov. 2016.

\bibitem{rob12}
A.~C. {Robin}, X.~{Luri}, C.~{Reyl{\'e}}, Y.~{Isasi}, E.~{Grux},
  S.~{Blanco-Cuaresma}, F.~{Arenou}, C.~{Babusiaux}, M.~{Belcheva},
  R.~{Drimmel}, C.~{Jordi}, A.~{Krone-Martins}, E.~{Masana}, J.~C. {Mauduit},
  F.~{Mignard}, N.~{Mowlavi}, B.~{Rocca-Volmerange}, P.~{Sartoretti},
  E.~{Slezak}, and A.~{Sozzetti}, ``{Gaia Universe model snapshot. A
  statistical analysis of the expected contents of the Gaia catalogue},'' {\em
  \aap}~{\bf 543}, p.~A100, July 2012.

\bibitem{duc13}
G.~{Duch{\^e}ne} and A.~{Kraus}, ``{Stellar Multiplicity},'' {\em \araa}~{\bf
  51}, pp.~269--310, Aug. 2013.

\bibitem{gra18}
{Gravity Collaboration}, E.~{Sturm}, J.~{Dexter}, O.~{Pfuhl}, M.~R. {Stock},
  R.~I. {Davies}, D.~{Lutz}, Y.~{Cl{\'e}net}, A.~{Eckart}, F.~{Eisenhauer},
  R.~{Genzel}, D.~{Gratadour}, S.~F. {H{\"o}nig}, M.~{Kishimoto}, S.~{Lacour},
  F.~{Millour}, H.~{Netzer}, G.~{Perrin}, B.~M. {Peterson}, P.~O. {Petrucci},
  D.~{Rouan}, I.~{Waisberg}, J.~{Woillez}, A.~{Amorim}, W.~{Brandner}, N.~M.
  {F{\"o}rster Schreiber}, P.~J.~V. {Garcia}, S.~{Gillessen}, T.~{Ott},
  T.~{Paumard}, K.~{Perraut}, S.~{Scheithauer}, C.~{Straubmeier}, L.~J.
  {Tacconi}, and F.~{Widmann}, ``{Spatially resolved rotation of the broad-line
  region of a quasar at sub-parsec scale},'' {\em \nat}~{\bf 563},
  pp.~657--660, Nov. 2018.

\bibitem{fei12}
G.~A. {Feiden} and B.~{Chaboyer}, ``{Reevaluating the Mass-Radius Relation for
  Low-mass, Main-sequence Stars},'' {\em \apj}~{\bf 757}, p.~42, Sept. 2012.

\bibitem{man15}
A.~W. {Mann}, G.~A. {Feiden}, E.~{Gaidos}, T.~{Boyajian}, and K.~{von Braun},
  ``{How to Constrain Your M Dwarf: Measuring Effective Temperature, Bolometric
  Luminosity, Mass, and Radius},'' {\em \apj}~{\bf 804}, p.~64, May 2015.

\bibitem{sta14}
K.~G. {Stassun}, G.~A. {Feiden}, and G.~{Torres}, ``{Empirical tests of
  pre-main-sequence stellar evolution models with eclipsing binaries},'' {\em
  \nar}~{\bf 60}, pp.~1--28, June 2014.

\bibitem{cha20}
W.~J. {Chaplin}, A.~M. {Serenelli}, A.~{Miglio}, T.~{Morel}, J.~T. {Mackereth},
  F.~{Vincenzo}, H.~{Kjeldsen}, S.~{Basu}, W.~H. {Ball}, A.~{Stokholm},
  K.~{Verma}, J.~R. {Mosumgaard}, V.~{Silva Aguirre}, A.~{Mazumdar},
  P.~{Ranadive}, H.~M. {Antia}, Y.~{Lebreton}, J.~{Ong}, T.~{Appourchaux},
  T.~R. {Bedding}, J.~{Christensen-Dalsgaard}, O.~{Creevey}, R.~A.
  {Garc{\'\i}a}, R.~{Handberg}, D.~{Huber}, S.~D. {Kawaler}, M.~N. {Lund},
  T.~S. {Metcalfe}, K.~G. {Stassun}, M.~{Bazot}, P.~G. {Beck}, K.~J. {Bell},
  M.~{Bergemann}, D.~L. {Buzasi}, O.~{Benomar}, D.~{Bossini}, L.~{Bugnet},
  T.~L. {Campante}, Z.~{\c{c}}. {Orhan}, E.~{Corsaro},
  L.~{Gonz{\'a}lez-Cuesta}, G.~R. {Davies}, M.~P. {Di Mauro}, R.~{Egeland},
  Y.~P. {Elsworth}, P.~{Gaulme}, H.~{Ghasemi}, Z.~{Guo}, O.~J. {Hall},
  A.~{Hasanzadeh}, S.~{Hekker}, R.~{Howe}, J.~M. {Jenkins}, A.~{Jim{\'e}nez},
  R.~{Kiefer}, J.~S. {Kuszlewicz}, T.~{Kallinger}, D.~W. {Latham}, M.~S.
  {Lundkvist}, S.~{Mathur}, J.~{Montalb{\'a}n}, B.~{Mosser}, A.~M. {Bed{\'o}n},
  M.~B. {Nielsen}, S.~{{\"O}rtel}, B.~M. {Rendle}, G.~R. {Ricker}, T.~S.
  {Rodrigues}, I.~W. {Roxburgh}, H.~{Safari}, M.~{Schofield}, S.~{Seager},
  B.~{Smalley}, D.~{Stello}, R.~{Szab{\'o}}, J.~{Tayar}, N.~{Theme{\ss}l},
  A.~E.~L. {Thomas}, R.~K. {Vanderspek}, W.~E. {van Rossem}, M.~{Vrard},
  A.~{Weiss}, T.~R. {White}, J.~N. {Winn}, and M.~{Y{\i}ld{\i}z}, ``{Age dating
  of an early Milky Way merger via asteroseismology of the naked-eye star
  {\ensuremath{\nu}} Indi},'' {\em Nature Astronomy}~{\bf 4}, pp.~382--389,
  Jan. 2020.

\bibitem{bro18}
K.~{Brogaard}, C.~J. {Hansen}, A.~{Miglio}, D.~{Slumstrup}, S.~{Frandsen},
  J.~{Jessen-Hansen}, M.~N. {Lund}, D.~{Bossini}, A.~{Thygesen}, G.~R.
  {Davies}, W.~J. {Chaplin}, T.~{Arentoft}, H.~{Bruntt}, F.~{Grundahl}, and
  R.~{Handberg}, ``{Establishing the accuracy of asteroseismic mass and radius
  estimates of giant stars - I. Three eclipsing systems at [Fe/H]
  {\ensuremath{\sim}} -0.3 and the need for a large high-precision sample},''
  {\em \mnras}~{\bf 476}, pp.~3729--3743, May 2018.

\bibitem{cam16}
T.~L. {Campante}, M.~{Schofield}, J.~S. {Kuszlewicz}, L.~{Bouma}, W.~J.
  {Chaplin}, D.~{Huber}, J.~{Christensen-Dalsgaard}, H.~{Kjeldsen},
  D.~{Bossini}, T.~S.~H. {North}, T.~{Appourchaux}, D.~W. {Latham},
  J.~{Pepper}, G.~R. {Ricker}, K.~G. {Stassun}, R.~{Vanderspek}, and J.~N.
  {Winn}, ``{The Asteroseismic Potential of TESS: Exoplanet-host Stars},'' {\em
  \apj}~{\bf 830}, p.~138, Oct. 2016.

\bibitem{mig17}
A.~{Miglio}, C.~{Chiappini}, B.~{Mosser}, G.~R. {Davies}, K.~{Freeman},
  L.~{Girardi}, P.~{Jofr{\'e}}, D.~{Kawata}, B.~M. {Rendle}, M.~{Valentini},
  L.~{Casagrande}, W.~J. {Chaplin}, G.~{Gilmore}, K.~{Hawkins}, B.~{Holl},
  T.~{Appourchaux}, K.~{Belkacem}, D.~{Bossini}, K.~{Brogaard}, M.~J. {Goupil},
  J.~{Montalb{\'a}n}, A.~{Noels}, F.~{Anders}, T.~{Rodrigues}, G.~{Piotto},
  D.~{Pollacco}, H.~{Rauer}, C.~{Allende Prieto}, P.~P. {Avelino},
  C.~{Babusiaux}, C.~{Barban}, B.~{Barbuy}, S.~{Basu}, F.~{Baudin},
  O.~{Benomar}, O.~{Bienaym{\'e}}, J.~{Binney}, J.~{Bland-Hawthorn},
  A.~{Bressan}, C.~{Cacciari}, T.~L. {Campante}, S.~{Cassisi},
  J.~{Christensen-Dalsgaard}, F.~{Combes}, O.~{Creevey}, M.~S. {Cunha}, R.~S.
  {Jong}, P.~{Laverny}, S.~{Degl'Innocenti}, S.~{Deheuvels}, {\'E}.~{Depagne},
  J.~{Ridder}, P.~{Di Matteo}, M.~P. {Di Mauro}, M.~A. {Dupret},
  P.~{Eggenberger}, Y.~{Elsworth}, B.~{Famaey}, S.~{Feltzing}, R.~A.
  {Garc{\'\i}a}, O.~{Gerhard}, B.~K. {Gibson}, L.~{Gizon}, M.~{Haywood},
  R.~{Handberg}, U.~{Heiter}, S.~{Hekker}, D.~{Huber}, R.~{Ibata}, D.~{Katz},
  S.~D. {Kawaler}, H.~{Kjeldsen}, D.~W. {Kurtz}, N.~{Lagarde}, Y.~{Lebreton},
  M.~N. {Lund}, S.~R. {Majewski}, P.~{Marigo}, M.~{Martig}, S.~{Mathur},
  I.~{Minchev}, T.~{Morel}, S.~{Ortolani}, M.~H. {Pinsonneault}, B.~{Plez},
  P.~G. {Prada Moroni}, D.~{Pricopi}, A.~{Recio-Blanco}, C.~{Reyl{\'e}},
  A.~{Robin}, I.~W. {Roxburgh}, M.~{Salaris}, B.~X. {Santiago}, R.~{Schiavon},
  A.~{Serenelli}, S.~{Sharma}, V.~{Silva Aguirre}, C.~{Soubiran},
  M.~{Steinmetz}, D.~{Stello}, K.~G. {Strassmeier}, P.~{Ventura}, R.~{Ventura},
  N.~A. {Walton}, and C.~C. {Worley}, ``{PLATO as it is : A legacy mission for
  Galactic archaeology},'' {\em Astronomische Nachrichten}~{\bf 338},
  pp.~644--661, July 2017.

\bibitem{que00}
D.~{Queloz}, A.~{Eggenberger}, M.~{Mayor}, C.~{Perrier}, J.~L. {Beuzit},
  D.~{Naef}, J.~P. {Sivan}, and S.~{Udry}, ``{Detection of a spectroscopic
  transit by the planet orbiting the star HD209458},'' {\em \aap}~{\bf 359},
  pp.~L13--L17, July 2000.

\bibitem{tri17}
A.~H.~M.~J. {Triaud}, D.~V. {Martin}, D.~{S{\'e}gransan}, B.~{Smalley},
  P.~F.~L. {Maxted}, D.~R. {Anderson}, F.~{Bouchy}, A.~{Collier Cameron},
  F.~{Faedi}, Y.~{G{\'o}mez Maqueo Chew}, L.~{Hebb}, C.~{Hellier},
  M.~{Marmier}, F.~{Pepe}, D.~{Pollacco}, D.~{Queloz}, S.~{Udry}, and
  R.~{West}, ``{The EBLM Project. IV. Spectroscopic orbits of over 100
  eclipsing M dwarfs masquerading as transiting hot Jupiters},'' {\em
  \aap}~{\bf 608}, p.~A129, Dec. 2017.

\bibitem{lai11}
D.~{Lai}, F.~{Foucart}, and D.~N.~C. {Lin}, ``{Evolution of spin direction of
  accreting magnetic protostars and spin-orbit misalignment in exoplanetary
  systems},'' {\em \mnras}~{\bf 412}, pp.~2790--2798, Apr. 2011.

\bibitem{alb22}
S.~H. {Albrecht}, R.~I. {Dawson}, and J.~N. {Winn}, ``{Stellar obliquities in
  exoplanetary systems},'' {\em arXiv e-prints} , p.~arXiv:2203.05460, Mar.
  2022.

\bibitem{kra20a}
S.~{Kraus}, J.-B. {Le Bouquin}, A.~{Kreplin}, C.~L. {Davies}, E.~{Hone}, J.~D.
  {Monnier}, T.~{Gardner}, G.~{Kennedy}, and S.~{Hinkley}, ``{Spin-Orbit
  Alignment of the {\ensuremath{\beta}} Pictoris Planetary System},'' {\em
  \apjl}~{\bf 897}, p.~L8, July 2020.

\bibitem{bat18}
M.~R. {Bate}, ``{On the diversity and statistical properties of protostellar
  discs},'' {\em \mnras}~{\bf 475}, pp.~5618--5658, Apr. 2018.

\bibitem{fab07}
D.~{Fabrycky} and S.~{Tremaine}, ``{Shrinking Binary and Planetary Orbits by
  Kozai Cycles with Tidal Friction},'' {\em \apj}~{\bf 669}, pp.~1298--1315,
  Nov. 2007.

\bibitem{gua04}
A.~{Gualandris}, S.~{Portegies Zwart}, and P.~P. {Eggleton}, ``{N-body
  simulations of stars escaping from the Orion nebula},'' {\em \mnras}~{\bf
  350}, pp.~615--626, May 2004.

\bibitem{leb09}
J.~B. {Le Bouquin}, O.~{Absil}, M.~{Benisty}, F.~{Massi}, A.~{M{\'e}rand}, and
  S.~{Stefl}, ``{The spin-orbit alignment of the Fomalhaut planetary system
  probed by optical long baseline interferometry},'' {\em \aap}~{\bf 498},
  pp.~L41--L44, May 2009.

\bibitem{lun14}
M.~N. {Lund}, M.~{Lundkvist}, V.~{Silva Aguirre}, G.~{Houdek}, L.~{Casagrande},
  V.~{Van Eylen}, T.~L. {Campante}, C.~{Karoff}, H.~{Kjeldsen}, S.~{Albrecht},
  W.~J. {Chaplin}, M.~B. {Nielsen}, P.~{Degroote}, G.~R. {Davies}, and
  R.~{Handberg}, ``{Asteroseismic inference on the spin-orbit misalignment and
  stellar parameters of HAT-P-7},'' {\em \aap}~{\bf 570}, p.~A54, Oct. 2014.

\bibitem{nic17}
C.~P. {Nicholls}, T.~{Lebzelter}, A.~{Smette}, B.~{Wolff}, H.~{Hartman}, H.~U.
  {K{\"a}ufl}, N.~{Przybilla}, S.~{Ramsay}, S.~{Uttenthaler}, G.~M. {Wahlgren},
  S.~{Bagnulo}, G.~A.~J. {Hussain}, M.~F. {Nieva}, U.~{Seemann}, and
  A.~{Seifahrt}, ``{CRIRES-POP: a library of high resolution spectra in the
  near-infrared. II. Data reduction and the spectrum of the K giant 10
  Leonis},'' {\em \aap}~{\bf 598}, p.~A79, Feb. 2017.

\bibitem{gra20}
{Gravity Collaboration}, R.~{Garcia Lopez}, A.~{Natta}, A.~{Caratti o Garatti},
  T.~P. {Ray}, R.~{Fedriani}, M.~{Koutoulaki}, L.~{Klarmann}, K.~{Perraut},
  J.~{Sanchez-Bermudez}, M.~{Benisty}, C.~{Dougados}, L.~{Labadie},
  W.~{Brandner}, P.~J.~V. {Garcia}, T.~{Henning}, P.~{Caselli}, G.~{Duvert},
  T.~{de Zeeuw}, R.~{Grellmann}, R.~{Abuter}, A.~{Amorim}, M.~{Baub{\"o}ck},
  J.~P. {Berger}, H.~{Bonnet}, A.~{Buron}, Y.~{Cl{\'e}net}, V.~{Coud{\'e} Du
  Foresto}, W.~{de Wit}, A.~{Eckart}, F.~{Eisenhauer}, M.~{Filho}, F.~{Gao},
  C.~E. {Garcia Dabo}, E.~{Gendron}, R.~{Genzel}, S.~{Gillessen}, M.~{Habibi},
  X.~{Haubois}, F.~{Haussmann}, S.~{Hippler}, Z.~{Hubert}, M.~{Horrobin},
  A.~{Jimenez Rosales}, L.~{Jocou}, P.~{Kervella}, J.~{Kolb}, S.~{Lacour},
  J.~B. {Le Bouquin}, P.~{L{\'e}na}, T.~{Ott}, T.~{Paumard}, G.~{Perrin},
  O.~{Pfuhl}, A.~{Ramirez}, C.~{Rau}, G.~{Rousset}, S.~{Scheithauer},
  J.~{Shangguan}, J.~{Stadler}, O.~{Straub}, C.~{Straubmeier}, E.~{Sturm},
  E.~{van Dishoeck}, F.~{Vincent}, S.~{von Fellenberg}, F.~{Widmann},
  E.~{Wieprecht}, M.~{Wiest}, E.~{Wiezorrek}, J.~{Woillez}, S.~{Yazici}, and
  G.~{Zins}, ``{A measure of the size of the magnetospheric accretion region in
  TW Hydrae},'' {\em \nat}~{\bf 584}, pp.~547--550, Aug. 2020.

\bibitem{hon17}
E.~{Hone}, S.~{Kraus}, A.~{Kreplin}, K.-H. {Hofmann}, G.~{Weigelt},
  T.~{Harries}, and J.~{Kluska}, ``{Gas dynamics in the inner few AU around the
  Herbig B[e] star MWC297. Indications of a disk wind from kinematic modeling
  and velocity-resolved interferometric imaging},'' {\em \aap}~{\bf 607},
  p.~A17, Oct. 2017.

\bibitem{wei16}
G.~{Weigelt}, K.~H. {Hofmann}, D.~{Schertl}, N.~{Clementel}, M.~F. {Corcoran},
  A.~{Damineli}, W.~J. {de Wit}, R.~{Grellmann}, J.~{Groh}, S.~{Guieu},
  T.~{Gull}, M.~{Heininger}, D.~J. {Hillier}, C.~A. {Hummel}, S.~{Kraus},
  T.~{Madura}, A.~{Mehner}, A.~{M{\'e}rand}, F.~{Millour}, A.~F.~J. {Moffat},
  K.~{Ohnaka}, F.~{Patru}, R.~G. {Petrov}, S.~{Rengaswamy}, N.~D. {Richardson},
  T.~{Rivinius}, M.~{Sch{\"o}ller}, M.~{Teodoro}, and M.~{Wittkowski},
  ``{VLTI-AMBER velocity-resolved aperture-synthesis imaging of
  {\ensuremath{\eta}} Carinae with a spectral resolution of 12 000. Studies of
  the primary star wind and innermost wind-wind collision zone},'' {\em
  \aap}~{\bf 594}, p.~A106, Oct. 2016.

\bibitem{gra17}
{Gravity Collaboration}, R.~{Abuter}, M.~{Accardo}, A.~{Amorim}, N.~{Anugu},
  G.~{{\'A}vila}, N.~{Azouaoui}, M.~{Benisty}, J.~P. {Berger}, N.~{Blind},
  H.~{Bonnet}, P.~{Bourget}, W.~{Brandner}, R.~{Brast}, A.~{Buron},
  L.~{Burtscher}, F.~{Cassaing}, F.~{Chapron}, {\'E}.~{Choquet},
  Y.~{Cl{\'e}net}, C.~{Collin}, V.~{Coud{\'e} Du Foresto}, W.~{de Wit}, P.~T.
  {de Zeeuw}, C.~{Deen}, F.~{Delplancke-Str{\"o}bele}, R.~{Dembet}, F.~{Derie},
  J.~{Dexter}, G.~{Duvert}, M.~{Ebert}, A.~{Eckart}, F.~{Eisenhauer},
  M.~{Esselborn}, P.~{F{\'e}dou}, G.~{Finger}, P.~{Garcia}, C.~E. {Garcia
  Dabo}, R.~{Garcia Lopez}, E.~{Gendron}, R.~{Genzel}, S.~{Gillessen},
  F.~{Gonte}, P.~{Gordo}, M.~{Grould}, U.~{Gr{\"o}zinger}, S.~{Guieu},
  P.~{Haguenauer}, O.~{Hans}, X.~{Haubois}, M.~{Haug}, F.~{Haussmann},
  T.~{Henning}, S.~{Hippler}, M.~{Horrobin}, A.~{Huber}, Z.~{Hubert},
  N.~{Hubin}, C.~A. {Hummel}, G.~{Jakob}, A.~{Janssen}, L.~{Jochum},
  L.~{Jocou}, A.~{Kaufer}, S.~{Kellner}, S.~{Kendrew}, L.~{Kern},
  P.~{Kervella}, M.~{Kiekebusch}, R.~{Klein}, Y.~{Kok}, J.~{Kolb}, M.~{Kulas},
  S.~{Lacour}, V.~{Lapeyr{\`e}re}, B.~{Lazareff}, J.~B. {Le Bouquin},
  P.~{L{\`e}na}, R.~{Lenzen}, S.~{L{\'e}v{\^e}que}, M.~{Lippa}, Y.~{Magnard},
  L.~{Mehrgan}, M.~{Mellein}, A.~{M{\'e}rand}, J.~{Moreno-Ventas}, T.~{Moulin},
  E.~{M{\"u}ller}, F.~{M{\"u}ller}, U.~{Neumann}, S.~{Oberti}, T.~{Ott},
  L.~{Pallanca}, J.~{Panduro}, L.~{Pasquini}, T.~{Paumard}, I.~{Percheron},
  K.~{Perraut}, G.~{Perrin}, A.~{Pfl{\"u}ger}, O.~{Pfuhl}, T.~{Phan Duc}, P.~M.
  {Plewa}, D.~{Popovic}, S.~{Rabien}, A.~{Ram{\'\i}rez}, J.~{Ramos}, C.~{Rau},
  M.~{Riquelme}, R.~R. {Rohloff}, G.~{Rousset}, J.~{Sanchez-Bermudez},
  S.~{Scheithauer}, M.~{Sch{\"o}ller}, N.~{Schuhler}, J.~{Spyromilio},
  C.~{Straubmeier}, E.~{Sturm}, M.~{Suarez}, K.~R.~W. {Tristram}, N.~{Ventura},
  F.~{Vincent}, I.~{Waisberg}, I.~{Wank}, J.~{Weber}, E.~{Wieprecht},
  M.~{Wiest}, E.~{Wiezorrek}, M.~{Wittkowski}, J.~{Woillez}, B.~{Wolff},
  S.~{Yazici}, D.~{Ziegler}, and G.~{Zins}, ``{First light for GRAVITY: Phase
  referencing optical interferometry for the Very Large Telescope
  Interferometer},'' {\em \aap}~{\bf 602}, p.~A94, June 2017.

\bibitem{alc14}
J.~M. {Alcal{\'a}}, A.~{Natta}, C.~F. {Manara}, L.~{Spezzi}, B.~{Stelzer},
  A.~{Frasca}, K.~{Biazzo}, E.~{Covino}, S.~{Randich}, E.~{Rigliaco},
  L.~{Testi}, F.~{Comer{\'o}n}, G.~{Cupani}, and V.~{D'Elia}, ``{X-shooter
  spectroscopy of young stellar objects. IV. Accretion in low-mass stars and
  substellar objects in Lupus},'' {\em \aap}~{\bf 561}, p.~A2, Jan. 2014.

\bibitem{kra08b}
S.~{Kraus}, K.~H. {Hofmann}, M.~{Benisty}, J.~P. {Berger}, O.~{Chesneau},
  A.~{Isella}, F.~{Malbet}, A.~{Meilland}, N.~{Nardetto}, A.~{Natta},
  T.~{Preibisch}, D.~{Schertl}, M.~{Smith}, P.~{Stee}, E.~{Tatulli},
  L.~{Testi}, and G.~{Weigelt}, ``{The origin of hydrogen line emission for
  five Herbig Ae/Be stars spatially resolved by VLTI/AMBER
  spectro-interferometry},'' {\em \aap}~{\bf 489}, pp.~1157--1173, Oct. 2008.

\bibitem{moe17}
M.~{Moe} and R.~{Di Stefano}, ``{Mind Your Ps and Qs: The Interrelation between
  Period (P) and Mass-ratio (Q) Distributions of Binary Stars},'' {\em
  \apjs}~{\bf 230}, p.~15, June 2017.

\bibitem{grap}
{GRAVITY+ consortium}, ``{GRAVITY+: Towards Faint Science, All Sky, High
  Contrast, Milli-Arcsecond Optical Interferometric Imaging - White Paper and
  proposal},'' 2020.

\bibitem{son19}
Y.-Y. {Songsheng}, J.-M. {Wang}, Y.-R. {Li}, and P.~{Du}, ``{Differential
  Interferometric Signatures of Close Binaries of Supermassive Black Holes in
  Active Galactic Nuclei},'' {\em \apj}~{\bf 881}, p.~140, Aug. 2019.

\bibitem{hon20}
E.~{Hone}, {\em {Resolving the gas distribution and kinematics in the inner
  regions of protoplanetary disks}}.
\newblock PhD thesis, University of Exeter, UK, Jan. 2020.

\bibitem{rom15}
M.~M. {Romanova} and S.~P. {Owocki}, ``{Accretion, Outflows, and Winds of
  Magnetized Stars},'' {\em \ssr}~{\bf 191}, pp.~339--389, Oct. 2015.

\bibitem{gra19}
{Gravity Collaboration}, S.~{Lacour}, M.~{Nowak}, J.~{Wang}, O.~{Pfuhl},
  F.~{Eisenhauer}, R.~{Abuter}, A.~{Amorim}, N.~{Anugu}, M.~{Benisty}, J.~P.
  {Berger}, H.~{Beust}, N.~{Blind}, M.~{Bonnefoy}, H.~{Bonnet}, P.~{Bourget},
  W.~{Brandner}, A.~{Buron}, C.~{Collin}, B.~{Charnay}, F.~{Chapron},
  Y.~{Cl{\'e}net}, V.~{Coud{\'e} Du Foresto}, P.~T. {de Zeeuw}, C.~{Deen},
  R.~{Dembet}, J.~{Dexter}, G.~{Duvert}, A.~{Eckart}, N.~M. {F{\"o}rster
  Schreiber}, P.~{F{\'e}dou}, P.~{Garcia}, R.~{Garcia Lopez}, F.~{Gao},
  E.~{Gendron}, R.~{Genzel}, S.~{Gillessen}, P.~{Gordo}, A.~{Greenbaum},
  M.~{Habibi}, X.~{Haubois}, F.~{Hau{\ss}mann}, T.~{Henning}, S.~{Hippler},
  M.~{Horrobin}, Z.~{Hubert}, A.~{Jimenez Rosales}, L.~{Jocou}, S.~{Kendrew},
  P.~{Kervella}, J.~{Kolb}, A.~M. {Lagrange}, V.~{Lapeyr{\`e}re}, J.~B. {Le
  Bouquin}, P.~{L{\'e}na}, M.~{Lippa}, R.~{Lenzen}, A.~L. {Maire},
  P.~{Molli{\`e}re}, T.~{Ott}, T.~{Paumard}, K.~{Perraut}, G.~{Perrin},
  L.~{Pueyo}, S.~{Rabien}, A.~{Ram{\'\i}rez}, C.~{Rau},
  G.~{Rodr{\'\i}guez-Coira}, G.~{Rousset}, J.~{Sanchez-Bermudez},
  S.~{Scheithauer}, N.~{Schuhler}, O.~{Straub}, C.~{Straubmeier}, E.~{Sturm},
  L.~J. {Tacconi}, F.~{Vincent}, E.~F. {van Dishoeck}, S.~{von Fellenberg},
  I.~{Wank}, I.~{Waisberg}, F.~{Widmann}, E.~{Wieprecht}, M.~{Wiest},
  E.~{Wiezorrek}, J.~{Woillez}, S.~{Yazici}, D.~{Ziegler}, and G.~{Zins},
  ``{First direct detection of an exoplanet by optical interferometry.
  Astrometry and K-band spectroscopy of HR 8799 e},'' {\em \aap}~{\bf 623},
  p.~L11, Mar. 2019.

\bibitem{ben21}
M.~{Benisty}, J.~{Bae}, S.~{Facchini}, M.~{Keppler}, R.~{Teague}, A.~{Isella},
  N.~T. {Kurtovic}, L.~M. {P{\'e}rez}, A.~{Sierra}, S.~M. {Andrews},
  J.~{Carpenter}, I.~{Czekala}, C.~{Dominik}, T.~{Henning}, F.~{Menard},
  P.~{Pinilla}, and A.~{Zurlo}, ``{A Circumplanetary Disk around PDS70c},''
  {\em \apjl}~{\bf 916}, p.~L2, July 2021.

\bibitem{haf19}
S.~Y. {Haffert}, A.~J. {Bohn}, J.~{de Boer}, I.~A.~G. {Snellen},
  J.~{Brinchmann}, J.~H. {Girard}, C.~U. {Keller}, and R.~{Bacon}, ``{Two
  accreting protoplanets around the young star PDS 70},'' {\em Nature
  Astronomy}~{\bf 3}, pp.~749--754, June 2019.

\bibitem{aoy20}
Y.~{Aoyama}, G.-D. {Marleau}, C.~{Mordasini}, and M.~{Ikoma}, ``{Spectral
  appearance of the planetary-surface accretion shock: Global spectra and
  hydrogen-line profiles and fluxes},'' {\em arXiv e-prints} ,
  p.~arXiv:2011.06608, Nov. 2020.

\bibitem{szu20}
J.~{Szul{\'a}gyi} and B.~{Ercolano}, ``{Hydrogen Recombination Line
  Luminosities and Variability from Forming Planets},'' {\em \apj}~{\bf 902},
  p.~126, Oct. 2020.

\bibitem{wan21}
J.~J. {Wang}, A.~{Vigan}, S.~{Lacour}, M.~{Nowak}, T.~{Stolker}, R.~J. {De
  Rosa}, S.~{Ginzburg}, P.~{Gao}, R.~{Abuter}, A.~{Amorim},
  R.~{Asensio-Torres}, M.~{Baub{\"o}ck}, M.~{Benisty}, J.~P. {Berger},
  H.~{Beust}, J.~L. {Beuzit}, S.~{Blunt}, A.~{Boccaletti}, A.~{Bohn},
  M.~{Bonnefoy}, H.~{Bonnet}, W.~{Brandner}, F.~{Cantalloube}, P.~{Caselli},
  B.~{Charnay}, G.~{Chauvin}, E.~{Choquet}, V.~{Christiaens}, Y.~{Cl{\'e}net},
  V.~{Coud{\'e} Du Foresto}, A.~{Cridland}, P.~T. {de Zeeuw}, R.~{Dembet},
  J.~{Dexter}, A.~{Drescher}, G.~{Duvert}, A.~{Eckart}, F.~{Eisenhauer},
  S.~{Facchini}, F.~{Gao}, P.~{Garcia}, R.~{Garcia Lopez}, T.~{Gardner},
  E.~{Gendron}, R.~{Genzel}, S.~{Gillessen}, J.~{Girard}, X.~{Haubois},
  G.~{Hei{\ss}el}, T.~{Henning}, S.~{Hinkley}, S.~{Hippler}, M.~{Horrobin},
  M.~{Houll{\'e}}, Z.~{Hubert}, A.~{Jim{\'e}nez-Rosales}, L.~{Jocou},
  J.~{Kammerer}, M.~{Keppler}, P.~{Kervella}, M.~{Meyer}, L.~{Kreidberg}, A.~M.
  {Lagrange}, V.~{Lapeyr{\`e}re}, J.~B. {Le Bouquin}, P.~{L{\'e}na}, D.~{Lutz},
  A.~L. {Maire}, F.~{M{\'e}nard}, A.~{M{\'e}rand}, P.~{Molli{\`e}re}, J.~D.
  {Monnier}, D.~{Mouillet}, A.~{M{\"u}ller}, E.~{Nasedkin}, T.~{Ott},
  G.~P.~P.~L. {Otten}, C.~{Paladini}, T.~{Paumard}, K.~{Perraut}, G.~{Perrin},
  O.~{Pfuhl}, L.~{Pueyo}, J.~{Rameau}, L.~{Rodet}, G.~{Rodr{\'\i}guez-Coira},
  G.~{Rousset}, S.~{Scheithauer}, J.~{Shangguan}, T.~{Shimizu}, J.~{Stadler},
  O.~{Straub}, C.~{Straubmeier}, E.~{Sturm}, L.~J. {Tacconi}, E.~F. {van
  Dishoeck}, F.~{Vincent}, S.~D. {von Fellenberg}, K.~{Ward-Duong},
  F.~{Widmann}, E.~{Wieprecht}, E.~{Wiezorrek}, J.~{Woillez}, and {Gravity
  Collaboration}, ``{Constraining the Nature of the PDS 70 Protoplanets with
  VLTI/GRAVITY},'' {\em \aj}~{\bf 161}, p.~148, Mar. 2021.

\bibitem{mar22}
G.~D. {Marleau}, Y.~{Aoyama}, R.~{Kuiper}, K.~{Follette}, N.~J. {Turner},
  G.~{Cugno}, C.~F. {Manara}, S.~Y. {Haffert}, D.~{Kitzmann}, S.~C.
  {Ringqvist}, K.~R. {Wagner}, R.~{van Boekel}, S.~{Sallum}, M.~{Janson},
  T.~O.~B. {Schmidt}, L.~{Venuti}, C.~{Lovis}, and C.~{Mordasini}, ``{Accreting
  protoplanets: Spectral signatures and magnitude of gas and dust extinction at
  H {\ensuremath{\alpha}}},'' {\em \aap}~{\bf 657}, p.~A38, Jan. 2022.

\bibitem{aoy21}
Y.~{Aoyama}, G.-D. {Marleau}, M.~{Ikoma}, and C.~{Mordasini}, ``{Comparison of
  Planetary H{\ensuremath{\alpha}}-emission Models: A New Correlation with
  Accretion Luminosity},'' {\em \apjl}~{\bf 917}, p.~L30, Aug. 2021.

\bibitem{now20}
{Gravity Collaboration}, M.~{Nowak}, S.~{Lacour}, P.~{Molli{\`e}re}, J.~{Wang},
  B.~{Charnay}, E.~F. {van Dishoeck}, R.~{Abuter}, A.~{Amorim}, J.~P. {Berger},
  H.~{Beust}, M.~{Bonnefoy}, H.~{Bonnet}, W.~{Brandner}, A.~{Buron},
  F.~{Cantalloube}, C.~{Collin}, F.~{Chapron}, Y.~{Cl{\'e}net}, V.~{Coud{\'e}
  Du Foresto}, P.~T. {de Zeeuw}, R.~{Dembet}, J.~{Dexter}, G.~{Duvert},
  A.~{Eckart}, F.~{Eisenhauer}, N.~M. {F{\"o}rster Schreiber}, P.~{F{\'e}dou},
  R.~{Garcia Lopez}, F.~{Gao}, E.~{Gendron}, R.~{Genzel}, S.~{Gillessen},
  F.~{Hau{\ss}mann}, T.~{Henning}, S.~{Hippler}, Z.~{Hubert}, L.~{Jocou},
  P.~{Kervella}, A.~M. {Lagrange}, V.~{Lapeyr{\`e}re}, J.~B. {Le Bouquin},
  P.~{L{\'e}na}, A.~L. {Maire}, T.~{Ott}, T.~{Paumard}, C.~{Paladini},
  K.~{Perraut}, G.~{Perrin}, L.~{Pueyo}, O.~{Pfuhl}, S.~{Rabien}, C.~{Rau},
  G.~{Rodr{\'\i}guez-Coira}, G.~{Rousset}, S.~{Scheithauer}, J.~{Shangguan},
  O.~{Straub}, C.~{Straubmeier}, E.~{Sturm}, L.~J. {Tacconi}, F.~{Vincent},
  F.~{Widmann}, E.~{Wieprecht}, E.~{Wiezorrek}, J.~{Woillez}, S.~{Yazici}, and
  D.~{Ziegler}, ``{Peering into the formation history of {\ensuremath{\beta}}
  Pictoris b with VLTI/GRAVITY long-baseline interferometry},'' {\em \aap}~{\bf
  633}, p.~A110, Jan. 2020.

\bibitem{mon18}
J.~{Monnier}, J.-B. {Le Bouquin}, N.~{Anugu}, S.~{Kraus}, B.~{Setterholm},
  J.~{Ennis}, C.~{Lanthermann}, L.~{Jocou}, and T.~{ten Brummelaar}, ``{MYSTIC:
  Michigan Young STar Imager at CHARA},'' in {\em Optical and Infrared
  Interferometry VI},  {\em \procspie}, July 2018.

\bibitem{por20}
E.~H. {Por} and S.~Y. {Haffert}, ``{The Single-mode Complex Amplitude
  Refinement (SCAR) coronagraph. I. Concept, theory, and design},'' {\em
  \aap}~{\bf 635}, p.~A55, Mar. 2020.

\bibitem{heimdallr_ireland}
M.~J. {Ireland}, D.~{Defr{\`e}re}, F.~{Martinache}, J.~D. {Monnier},
  B.~{Norris}, P.~{Tuthill}, and J.~{Woillez}, ``{Image-plane fringe tracker
  for adaptive-optics assisted long baseline interferometry},'' in {\em Optical
  and Infrared Interferometry and Imaging VI},  M.~J. {Creech-Eakman}, P.~G.
  {Tuthill}, and A.~{M{\'e}rand}, eds., {\em Society of Photo-Optical
  Instrumentation Engineers (SPIE) Conference Series} {\bf 10701}, p.~1070111,
  July 2018.

\bibitem{asgard_martinod}
M.-A. {Martinod}, A.~{Bigioli}, J.~{Bryant}, S.~{Chhabra},
  B.~{Courtney-Barrer}, F.~{Crous}, N.~{Cvetojevic}, C.~{Dandumont},
  G.~{Defr\`ere}, D.~{Garreau}, M.~{Ireland}, T.~{Kraus}, T.~{Lagadec},
  R.~{Laugier}, F.~{Martinache}, D.~{Mortimer}, B.~{Norris}, G.~{Robertson},
  A.~{Taras}, and P.~G. {Tuthill}, ``{High-angular and high-contrast VLTI
  observations from J to M band with the Asgard instrumental suite},'' in {\em
  Optical and Infrared Interferometry and Imaging VIII},  {\em Society of
  Photo-Optical Instrumentation Engineers (SPIE) Conference Series} {\bf
  12183}, pp.~12183--36, July 2022.

\bibitem{fin16}
G.~{Finger}, I.~{Baker}, D.~{Alvarez}, C.~{Dupuy}, D.~{Ives}, M.~{Meyer},
  L.~{Mehrgan}, J.~{Stegmeier}, and H.~J. {Weller}, ``{Sub-electron read noise
  and millisecond full-frame readout with the near infrared eAPD array
  SAPHIRA},'' in {\em Adaptive Optics Systems V},  E.~{Marchetti}, L.~M.
  {Close}, and J.-P. {V{\'e}ran}, eds., {\em Society of Photo-Optical
  Instrumentation Engineers (SPIE) Conference Series} {\bf 9909}, p.~990912,
  July 2016.

\bibitem{fea22}
P.~{Feautrier} and J.-L. {Gach}, ``{Last performances improvement of the C-RED
  One camera using the 320x256 e-APD infrared Saphira detector},'' {\em arXiv
  e-prints} , p.~arXiv:2208.00377, July 2022.

\bibitem{bifrost_mortimer}
D.~{Mortimer}, S.~{Kraus}, J.~D. {Monnier}, J.-B. {Le Bouquin}, N.~{Anugu}, and
  S.~{Chhabra}, ``{Beam combiner for the Asgard/BIFROST instrument},'' in {\em
  Optical and Infrared Interferometry and Imaging VIII},  {\em Society of
  Photo-Optical Instrumentation Engineers (SPIE) Conference Series} {\bf
  12183}, pp.~12183--68, July 2022.

\bibitem{laz12}
B.~{Lazareff}, J.~B. {Le Bouquin}, and J.~P. {Berger}, ``{A novel technique to
  control differential birefringence in optical interferometers. Demonstration
  on the PIONIER-VLTI instrument},'' {\em \aap}~{\bf 543}, p.~A31, July 2012.

\bibitem{bifrost_chhabra}
S.~{Chhabra}, M.~{Frangiamore}, S.~{Kraus}, A.~{Bianco}, F.~{Garzon}, J.~D.
  {Monnier}, and D.~{Mortimer}, ``{Spectrograph design for the Asgard/BIFROST
  spectro-interferometric instrument for the VLTI},'' in {\em Optical and
  Infrared Interferometry and Imaging VIII},  {\em Society of Photo-Optical
  Instrumentation Engineers (SPIE) Conference Series} {\bf 12183},
  pp.~12183--22, July 2022.

\end{thebibliography}
\bibliographystyle{spiebib}   

\end{document}